\documentclass{iopart}

\usepackage{graphicx}
\begin{document}

\review{Transition between nuclear and quark-gluon descriptions of hadrons and light nuclei}

\author{R J Holt$^1$, R Gilman$^2$}
\address{$^1$ Physics Division, Argonne National Laboratory, Argonne, Illinois, 60439, USA}
\ead{holt@anl.gov}
\address{$^2$Department of Physics, Rutgers University, Piscataway, New Jersey, 08854, USA}
\ead{gilman@jlab.org}

\begin{abstract}
We provide a perspective on studies aimed at observing the transition
between hadronic and quark-gluonic descriptions of reactions involving
light nuclei.  
We begin by summarizing the results for relatively simple reactions 
such as the pion form factor and the neutral pion transition form
factor as well as that for the nucleon and end with exclusive 
photoreactions in our simplest nuclei.  A particular focus will be on 
reactions involving the deuteron.  It is noted that a firm
understanding of these issues is essential for unravelling important 
structure information from processes such as deeply virtual Compton 
scattering as well as deeply virtual meson production.  
The connection to exotic phenomena such as color transparency 
will be discussed.  
A number of outstanding challenges will require new experiments at 
modern facilities on the horizon as well as further theoretical developments.
\end{abstract}

\pacs{25.20.-x, 25.30.-c, 14.20.Dh}
\maketitle

\section{INTRODUCTION}
\label{sec:intro}

\begin{figure}[ht]
\begin{center}
\includegraphics[width=4.0 in, angle=0]{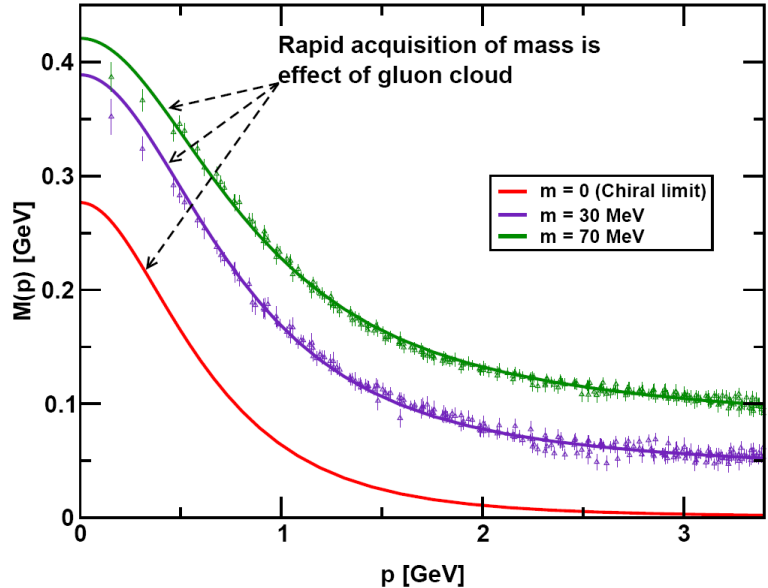}
\caption{ Dyson-Schwinger equation calculations of quark mass as a function
of the magnitude of the dressed quark's four-momentum for several current quark masses, as denoted by $m$. The points with error bars are from lattice QCD calculations.  The solid red curve is the result for a vanishing current quark mass, the chiral limit. 
[Reprinted from \cite{Roberts:2007ji}, Copyright (2007) with permission from Elsevier.]
\label{fig:dsmass}}
\end{center}
\end{figure}

One of the central goals of nuclear physics is the description of hadrons and nuclei at a truly fundamental level. While quantum chromodynamics (QCD) is the theory of the strong interaction, making use of this theory is one of the most challenging endeavors in science.  The problem is that non-perturbative methods must be used to describe the real world.  Observables are controlled by two emergent phenomena: confinement and dynamical chiral symmetry breaking (DCSB).  Dynamic chiral symmetry breaking is responsible for more than 98\% of the visible mass in the Universe.  The effect of DCSB has been studied through lattice calculations \cite{Bowman:2005vx}, the Dyson-Schwinger equation (DSE) approach \cite{Bhagwat:2003vw,Bhagwat:2006tu}, as well as instanton models \cite{Diakonov:1998rk}.  The results of these calculations are shown in Figure~\ref{fig:dsmass}. In this figure the mass of the quark is plotted as a function of the magnitude of the dressed quark's four-momentum.  Clearly as the quark momentum increases to 2\ GeV and beyond, the quark mass has fallen rapidly from its constituent quark mass to nearly its current quark mass.  Even under the assumption of perfect chiral symmetry, i.e.\ a vanishing quark mass as given by the solid red line in the figure, the quark mass evolves to essentially the constituent quark mass at low momentum.  In future experiments, it will be interesting to determine how this rapid change in quark mass can affect high-energy nuclear reactions.  The interesting regions will be reactions in kinematic regimes where the quark mass function changes rapidly.

One approach to our understanding of hadrons at this fundamental level is to determine the role of the quarks and gluons in hadronic and nuclear reactions. In particular, determining whether there is a clean transition from hadronic to quark-gluon degrees of freedom has been an important pursuit both experimentally and theoretically.  Historically, the constituent counting rule \cite{Brodsky:1974vy,Brodsky:1973kr,Matveev:1972gb}, hadron helicity conservation \cite{Brodsky:1981kj} and colour transparency effects \cite{Cosyn:2007er} have often been cited as evidence for the underlying quark degrees of freedom in reactions. The constituent counting rule states that the cross section, $d\sigma/dt$, should have a simple power law behaviour based on the number of constituents, $n$, involved in the process:  $d\sigma/dt \sim s^{2-n}$ where $s$ and $t$ are the usual Mandelstam variables.  Many experimental studies (See \cite{White:1994tj} for an example.) of exclusive reactions at high energies are consistent with the constituent counting rules.  It is believed that these effects should become manifest when perturbative QCD (pQCD) is valid.  In recent years, understanding exactly where pQCD and non-perturbative QCD are dominant has become important for studies of structure functions, the generalized parton distribution functions (GPDs), which provide information on quark position-momentum correlations, in particular.  For example, the exclusive processes of deeply virtual Compton scattering and deeply virtual meson production have been put forward as reactions necessary to isolate features of the GPDs, and depend on the process being factorizable into a hard 
production process and soft hadronic structure.

Intertwined with the idea of a quark-hadron transition is the idea of duality.
Bloom and Gilman \cite{Bloom:1970xb} introduced the idea, finding that an average over the resonance
region in inelastic electron scattering is equivalent to the scaling curve
in deep-inelastic scattering, and thus to quark behaviour.
Duality has been a ongoing topic of a number of experimental and theoretical investigations,
including extension from inclusive electron scattering to a variety of other reactions -- 
for an extensive review, see \cite{Melnitchouk:2005zr}.
A recent example of an experimental investigation \cite{Asaturyan:2011mq} of duality is in  
semi-inclusive pion production reactions.
While the idea of duality in the case of the nucleon is generally accepted, there is controversy over whether hadronic and
quark degrees of freedom are equivalent when considering the $NN$ force and nuclear structure,
as will be discussed in Section~\ref{sec:quarkvhadronic} below.

In this report, we will present highlights from the vast body of data aimed at discovering the transition from the hadronic picture which is well accepted at low energy and the QCD picture which is the theory of the strong interaction. The evidence is overwhelming that pQCD scaling is not achieved, except in the simplest systems, in exclusive reactions at contemporary kinematics.  Here, we focus on the form factors and transition form factors of the pion, nucleon and deuteron as well as photodisintegration of the nucleon, deuteron and $^3$He. In addition, we will review the evidence for the colour transparency effect which is believed to be a necessary precursor for factorization in semi-exclusive reactions.  In particular, we discuss results for high-energy exclusive reactions from  Stanford Linear Accelerator Center (SLAC), Jefferson Lab (JLab), Fermi National Accelerator Laboratory (FNAL) and Brookhaven National Laboratory (BNL). 
On the theoretical side the challenge is to calculate the kinematic dependences of the form factors, transition form factors, and reaction cross sections for simple systems.  Here we summarize the contemporary issues and approaches in the field.

\section{Quark-gluon vs. hadronic descriptions at low energy}
\label{sec:quarkvhadronic}

Quarks and gluons, the degrees of freedom of QCD, are confined within 
hadrons, the degrees of freedom that are detected by experiments.
Thus it seems obvious that in principle equivalent descriptions can be 
formulated in terms either of quark and gluon or of hadronic basis
states.
This viewpoint has however been challenged by some theorists,
since the early days of QCD and quark theories \cite{Isgur:2000pt53}.
Here we will give examples of three such arguments.
However this argument is resolved, it remains a practical question
whether it is possible to formulate a satisfactory theoretical
description with either set or both sets of basis states.

From the quark model point of view, it should be pointed out
that six-quark systems having the same quantum numbers as
baryon-baryon systems will in part have configurations that
do not break down into individual baryon quantum numbers
\cite{Harvey1981301,Harvey1981326,Brodsky:1983vf}.
The deuteron-like 6-quark wave function has the form
\begin{equation}
\psi = \sqrt{1/9} \, \psi_{NN} + \sqrt{4/45} \, \psi_{\Delta \Delta} + \sqrt{4/5} \, \psi_{CC},
\end{equation}
where the final $CC$ component is a non-baryonic hidden-colour component --
two three-quark systems, each with net colour, that add to a colourless deuteron.
The argument is that the hidden-colour component of the wave function
cannot be represented by colour-less hadrons.
However, a nonrelativistic constituent quark model calculation 
\cite{PhysRevD.29.952} found that there
is a strong dynamical clustering of the six-quark system into an
$NN$ configuration, with a strong repulsive core to the $NN$
interaction. This suggests that the hidden-colour component of the
$NN$ wave function is strongly suppressed for low-energy phenomena.

A second argument arises from the quark-meson coupling model
applied to nucleons in nuclei \cite{Guichon:2004xg}. In this viewpoint
it is unsurprising to find that nucleon structure is modified by the nucleon
being placed in a strong external field. Since the model leads to an effective
interaction in nuclei that agrees well with the phenomenological Skyrme force,
it supports the idea that one should think of nuclei as made up of quasi-particle
nucleons, as opposed to free nucleons.

An additional argument arises from a consideration of confinement
\cite{Ralston:2008sm}. Ralston argues that hadrons are incomplete
to describe their own interactions, when colour is exchanged. 
The system cannot be required to be colourless at all times, so
``{\em there is not supposed to be a local effective hadronic 
theory of any kind representing QCD.}'''

If these objections are valid, one might view lattice QCD or a
Dyson-Schwinger approach as the 
only theoretically acceptable solutions at present to low-energy QCD.
However, lattice QCD remains limited by computational capabilities,
with only some initial steps taken in exploring the $NN$ force.
Thus, even if these arguments are valid,  we anticipate that
QCD inspired effective hadronic field theories will remain a basis 
for our understanding low-energy QCD for many years.

\subsection{Effective field theories}
\label{sec:nopieft}

A number of theories related to QCD have been
developed to describe nonperturbative, low-energy
phenomena. Here we briefly describe 
Skyrme theory,
pionless effective field theory, EFT($\not\!\! \pi$), and 
chiral perturbation theory, $\chi$PT.

Skyrme theory \cite{Skyrme:1961vq} treats baryons as
topological solitons of an effective pion theory, justified 
by the large $N_c$ limit in which QCD becomes a theory 
of mesons \cite{'tHooft:1974hx}.
The theory has been applied to baryons, $NN$ interactions
\cite{Eisenberg1996321}, the structure of the deuteron -- see 
\cite{0034-4885-53-9-001} for a review of and references to earlier work, 
and more recently even to $\alpha$ particles \cite{PhysRevC.74.025203} 
and neutron stars \cite{PhysRevC.81.035203}.
Predictions tend to be qualitatively rather than quantitatively correct.

Modern EFT($\not\!\! \pi$) was developed first by Weinberg \cite{Weinberg1990288}.
The idea is that the physics at lower momentum than a scale
$m_{\pi}$ can be described with an expansion in powers of $p/m_{\pi}$
that reflects all desired symmetries. There remain issues and subtleties
with implementing the theory --
see \cite{RevModPhys.81.1773,Beane2002377,Gilman:2001yh} 
for further discussion.
In EFT($\not\!\! \pi$), $NN$ interactions arise from contact terms. 
An example of the structure of the deuteron in EFT($\not\!\! \pi$) is
\cite{Phillips2000209}.
The calculation quantitatively describes the deuteron form factors only 
up to $Q^2$ $\approx$ $m_{\pi}^2$, about as expected.
A strength of this approach is that the well-known issue of getting
the deuteron quadrupole moment correct is solved by fixing the constant of
a short-distance term involving a four-nucleon, one-photon contact term.

\begin{figure}[ht]
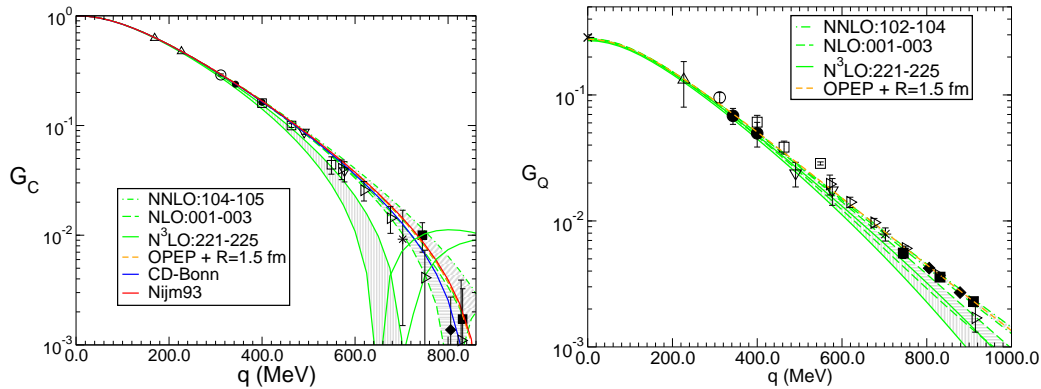
\vspace*{2mm}
\begin{center}
\includegraphics[height=2.0 in, angle=0]{figure_2a.eps}\hspace*{5mm}
\includegraphics[height=2.0 in, angle=0]{figure_2b.eps}
\caption{The charge (left) and quadrupole (right) form factors
of the deuteron in $\chi$PT.
[Reprinted from \cite{0954-3899-34-2-015}, Copyright (2007) with
permission from the author.]
\label{fig:chiptgcgq}}
\end{center}
\end{figure}

Using an expansion scale of $m_{\rho}$ adds pions to the EFT,
leading to $\chi$PT.
Calculations of the $NN$ force are now up to fourth order,
and describe $NN$ phase shifts well up to 250 MeV.
Earlier more qualitative predictions of the deuteron 
electromagnetic form factors, such as \cite{Walzl:2001vb},
have led to excellent quantitative predictions 
\cite{0954-3899-34-2-015} up to about $Q \approx m_{\rho}$,
as shown in Figure~\ref{fig:chiptgcgq}.
See Secs.~\ref{sec:deuterontheory} and \ref{sec:qgdeuteron} for 
discussion of the deuteron structure at higher $Q^2$.

\subsection{The issue of medium modifications}

At the beginning of this section we discussed the issue of hadronic
vs.\ quark-gluon theories.
When nucleons within nuclei are studied, the question arises
whether the properties of the nucleon are changed.
One viewpoint is that when a composite quark system, the nucleon,
is subjected to the strong external nuclear force, the properties of 
the system are modified.
The alternate viewpoint is that we have a many-body system of 
interacting hadrons, which can be described in terms of the properties
and interactions of the free hadrons.
These two viewpoints are related to the degrees of freedom used, and
might ultimately be different ways of looking at the same physics,
leading to equivalent predictions.
Even if the theories are not in principle equivalent, since hadronic
theories are based on the measured $NN$ force, any quark effects 
may be in part effectively accommodated by the hadronic theory.
In practice, the issue is whether observables are more simply
predicted from theories that incorporate quark-model inspired
medium modifications,  or whether observables are well understood
from hadronic theories without medium modifications.
Experimentally, this issue has been addressed by experiments concerning
the Coulomb sum rule, 
quasi-free electron scattering, 
polarization transfer to nucleons in nuclei, and
deep inelastic scattering on nuclei and the EMC effect.

\subsubsection{Coulomb sum rule}

Inclusive $(e,e^{\prime})$ scattering can be described as a sum
of two response functions, the transverse and longitudinal response functions
$R_T(\vec{q},\omega)$  and $R_L(\vec{q},\omega)$, respectively.
Here $\vec{q}$ and $\omega$ are the momentum and energy transfer.
The transverse (longitudinal) function $R_T$ ($R_L$) corresponds to 
virtual photons with transverse (longitudinal) electromagnetic fields 
like (unlike) the real photon, and reflects the magnetic (electric) structure
of the target. Following \cite{Benhar:2006wy}, the Coulomb sum rule 
can be defined as
\begin{equation}
S_L(\vec{q}) = {1 \over Z} \int_{\omega_0}^{\infty} 
{{R_L(\vec{q},\omega)}\over{\tilde{G}_E^2}} d\omega,
\end{equation}
where $\tilde{G}_E^2 = G_{Ep}^2 + N/Z \, G_{En}^2$ and
$\omega_0$ is the inelastic threshold.
Ignoring the neutron contributions, the integral in $S_L$ may
be thought of as counting the number of protons in the nucleus.
At low $\vec{q}$, below a few hundred MeV/$c$, 
nucleon correlations reduce the sum rule below unity, 
but it is believed that by about 500 MeV/$c$, deviations
of the sum rule from unity would be indicative of medium
modifications.
The experimental status of the Coulomb sum rule might
be regarded as not yet clear, due to conflicting analyses of the
world data - see \cite{Benhar:2006wy} for a discussion.
While recent theoretical work \cite{Wallace:2008ev} on 
Coulomb corrections, a major issue in the analyses, appears 
to support the idea that the Coulomb sum rule is quenched,
the uncertainties are not sufficient for a definite conclusion.
The situation should be improved in the near future due
to a recent JLab experiment \cite{Meziani:2005e0}.

\subsubsection{Quasifree electron scattering}

In the impulse approximation, the shape of the quasifree
scattering peak reflects the momentum distribution of nucleons
in nuclei, while its magnitude reflects the nucleon form factors.
Thus cross sections from different kinematics, and even from different
nuclei, can be checked for consistency with free nucleon form factors.
This is most often performed with cross sections rescaled by a scaling
function to follow a universal curve. Most familiar is probably
$y$ scaling, but there is also $\xi$ scaling, or superscaling 
with $\psi^{\prime}$. As discussed in \cite{Benhar:2006wy},
this technique has largely been used to set limits on medium
modifications, sensitive mostly to the magnetic form factor
of below $\approx$3\%.

\subsubsection{Polarization transfer to nucleons in nuclei}

\begin{figure}[ht]
\begin{center}
\includegraphics[height=2.0 in, angle=0]{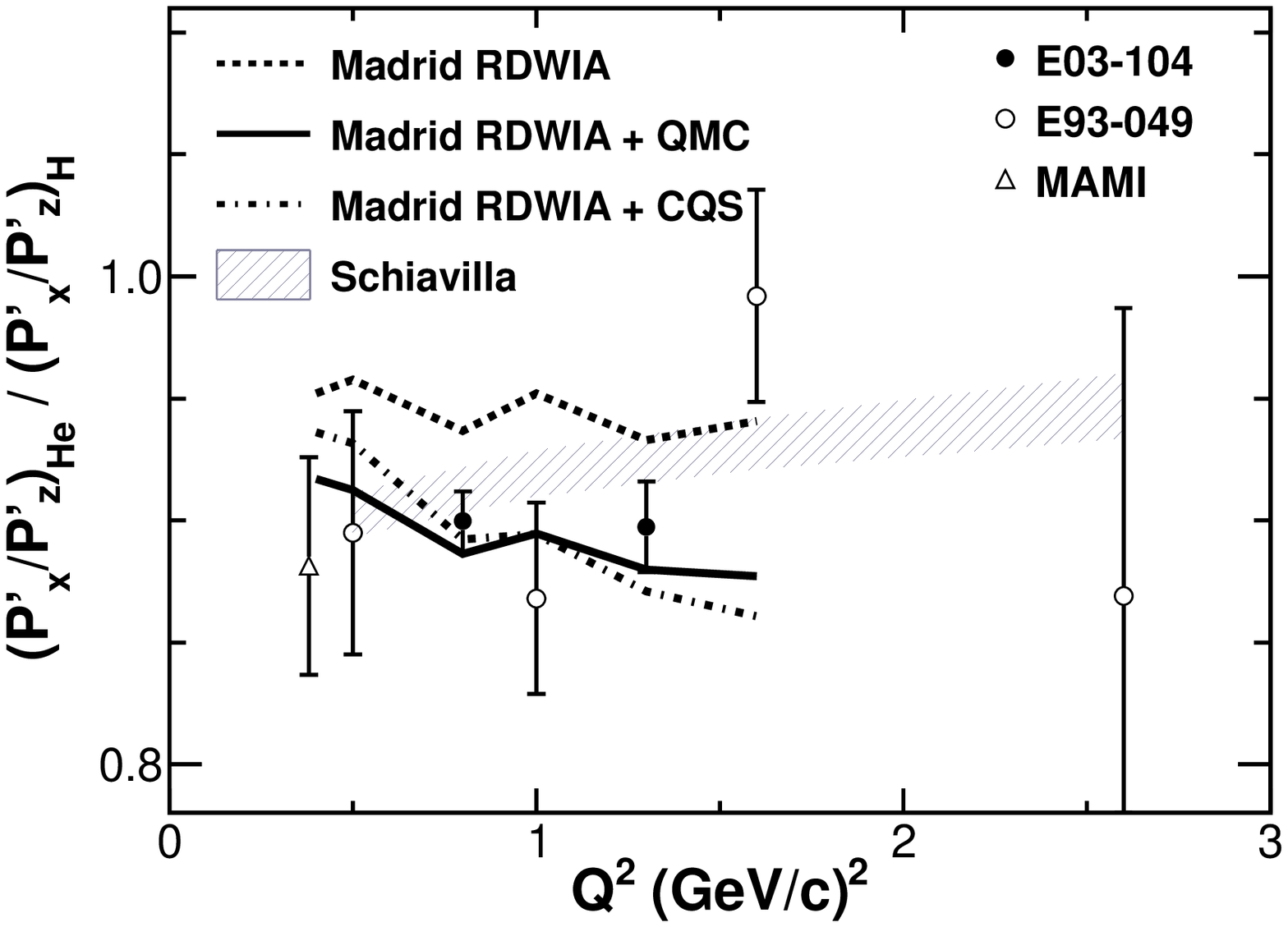}
\includegraphics[height=2.0 in, angle=0]{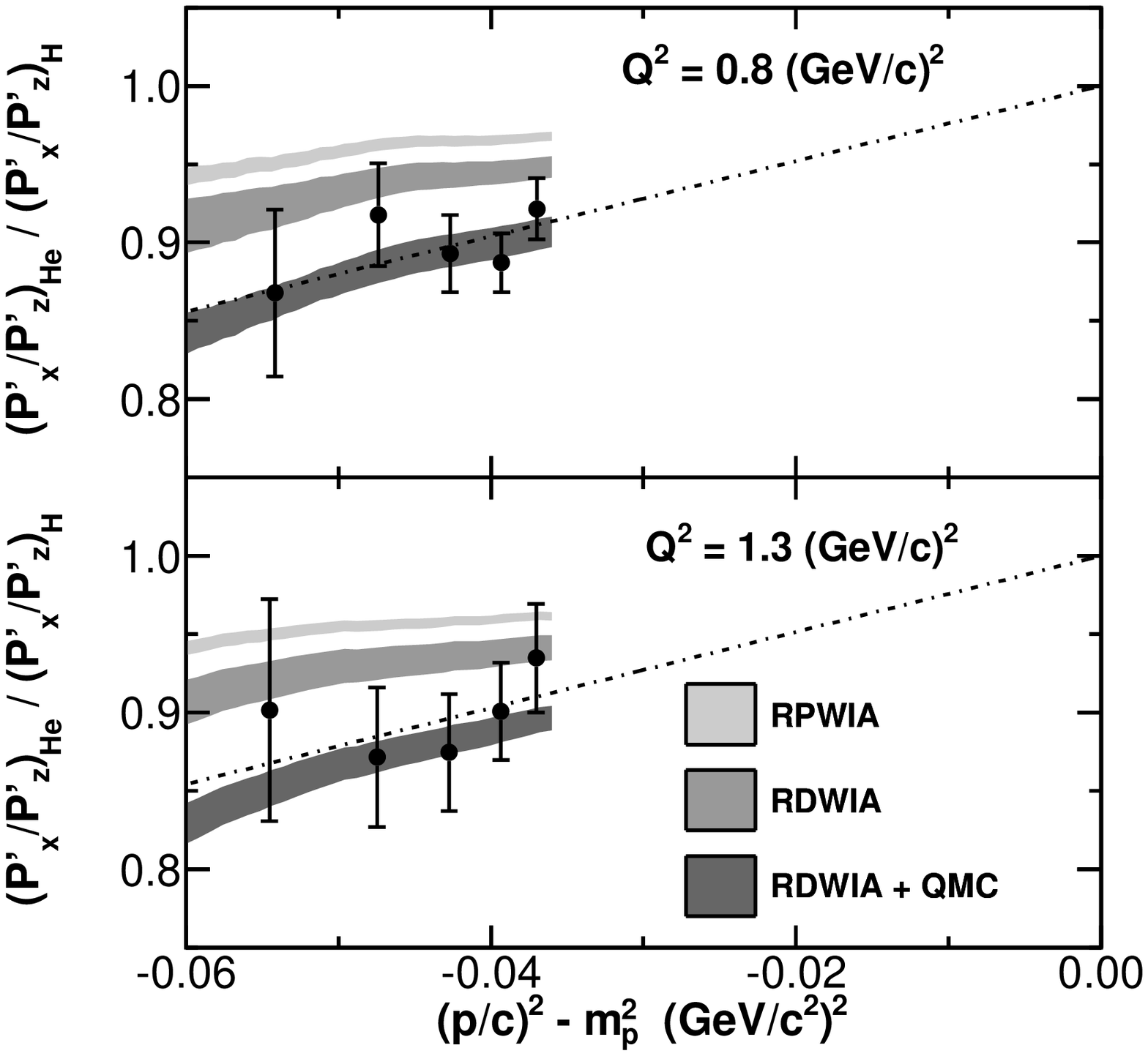}
\caption{The ratio of transverse to longitudinal polarization components
of a proton ejected from $^4$He compared to a free proton.
The left panel is integrated over the full acceptance, while the right
panel is for each of the E03-104 $Q^2$ points of \cite{Paolone:2010qc}
as a function of the initial state proton virtuality.
The Mainz point is from \cite{Dieterich:2000mu},
the E93-049 points are from \cite{Strauch:2002wu}.
Curves are described in the text.
Adapted from \cite{Paolone:2010qc}.
\label{fig:pffrat}}
\end{center}
\end{figure}

The $\vec{e} p \to e^{\prime} \vec{p}$ polarization transfer
reaction determines the proton form factor ratio through
\begin{equation}
{G_E \over G_M} = -{{E+E^{\prime}}\over 2M}\cot{\theta \over 2} \, {P_x \over P_z},
\end{equation}
where $P_{x,z}$ are polarization components of the final state proton,
$E$ ($E^{\prime}$) is the initial (final) state electron energy,
$M$ is the nucleon mass, and
$\theta$ is the electron scattering angle.
For protons in nuclei, the same ratio can be determined, although
the identification of this ratio with an in-medium form factor
ratio is suspect at best; formally there are 6 half-off-shell proton form factors.
The most recent experimental work, \cite{Paolone:2010qc},
reaffirmed with improved uncertainties that the proton polarization ratio
is reduced by about 10\% for protons ejected from $^4$He -- 
see Figure~\ref{fig:pffrat}.

This reduction in the ratio has been explained by two calculations.
First, calculations by the Madrid group \cite{Udias:1999tm} are unable to
reproduce the ratio without including medium modified nucleon form factors.
The quark meson coupling (QMC) modifications in Figure~\ref{fig:pffrat}
are from \cite{Lu:1997mu}, while the chiral quark soliton (CQS) modifications 
are from \cite{Smith:2004dn};
both models lead to similar results.
The QMC model of the nucleon uses constituent quarks confined in a
nucleon, with nucleons interacting through pions exchanged between quarks.
The CQS model of the nucleon is based on instantons in large $N_c$ QCD
and dynamical chiral symmetry breaking, and includes sea quarks absent
in the QMC approach.
The validity of the idea of medium modifications is supported
by a suggestion from \cite{CiofidegliAtti:2007vx}, that medium
modifications for low momentum should increase with the nucleon
virtuality;
the right panel of Figure~\ref{fig:pffrat} intriguingly shows such an effect.

Secondly, a conventional nuclear physics explanation is given by \cite{Schiavilla:2004xa}
in a much more detailed calculation that includes meson-exchange currents, 
tensor correlations, and spin-dependent and independent charge-exchange
final state interactions.
While the calculation of \cite{Udias:1999tm} arguably is too
simplistic, the calculation
of \cite{Schiavilla:2004xa}
can be criticized as not entirely constrained by data from other reactions.
The induced polarization in $^4$He$(e,e^{\prime}p)^3$H {\em suggests}
that the final-state interactions in \cite{Schiavilla:2004xa} are too strong,
but the result is not definitive.

Thus, the correct interpretation of the polarization transfer 
reactions appears inconclusive.
The next step in resolving this issue will likely come from the
interesting theoretical result of \cite{Cloet:2009tx}.
A model-independent prediction is that, while the form factor
ratio in the proton is expected to decrease, the form factor ratio in the
neutron is expected to increase.
As an experiment at JLab in the 12-GeV era appears unfeasible due
to the high beam energies, an experiment is being developed for MAMI
at Mainz.

\subsubsection{EMC effect}

The origins of the EMC effect \cite{Aubert:1983xm}, the depletion of 
quark distributions in nuclei at moderate Bjorken $x$, remain
ambiguous nearly 30 years after the effect was first observed.
A number of experiments have confirmed the 
depletion seen in the EMC results, and
it is generally accepted that the explanation must lie in a
modification of the quark distribution of nucleons.
In \cite{PhysRevC.65.015211}, it was argued that the EMC effect could not
be explained on the basis of a nucleons-only model of the nucleus,
and that constraints on antiquarks in nuclei are inconsistent
with explaining the EMC effect within a nucleon $+$ meson
model of nuclei. There have been interesting attempts to explain the
EMC effect based on many-body theory \cite{Benhar:1997vy,Benhar:2000wi}.  
Although these approaches have not been ruled out, there is
substantially more work necessary to successfully describe the 
effect without resorting to partonic descriptions.

Deep-inelastic scattering from light nuclei is a particularly powerful approach to study 
medium modifications since realistic nuclear calculations can be 
performed and since Coulomb effects \cite{Frankfurt:2010cb} are minimized.
Recent measurements in light nuclei \cite{Seely:2009gt} appear
to show that the EMC effect correlates more with local 
density, for example alpha clusters in $^9$Be, than with average nuclear density. 
Also, a recent analysis \cite{Weinstein:2010rt} indicates a
correlation between the strength of the EMC effect and the
strength of short-range correlations in nuclei.
However, these clues do not uniquely identify the underlying dynamics.
New measurements that will provide helpful information include
improved measurements of the EMC effect in the Drell-Yan process \cite{Isenhower:2001zz},
studies of quark-flavour dependence in the EMC effect, a measurement
of the EMC effect in the triton \cite{Petratos:2006},
and a possible spin-dependence in the EMC effect \cite{PhysRevLett.95.052302}.

\section{Transition from hadronic to quark-gluon degrees of freedom}
\subsection{The Pion}
\subsubsection{The Pion Elastic Form Factor}
\label{sec:pion}

\begin{figure}[ht]
\begin{center}
\includegraphics[width=5.5 in, angle=0]{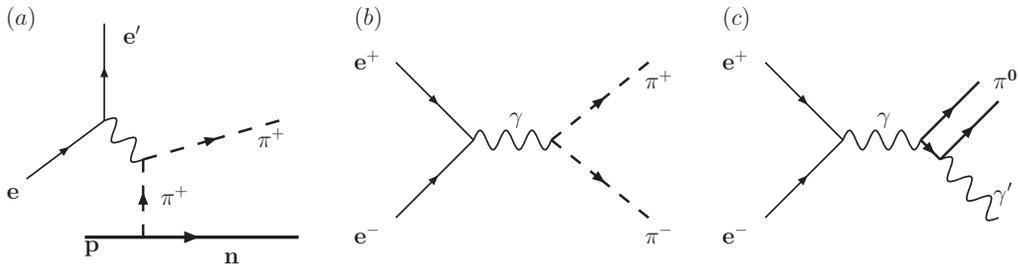}
\caption{Diagrams of (a) electron scattering from a virtual pion in a proton, (b) the $e^+e^- \rightarrow \gamma^* \rightarrow \pi^+ \pi^-$ reaction, and (c) the $e^+e^- \rightarrow \gamma^* \rightarrow \pi^\circ e^+ e^-$ reaction.
\label{fig:pifeyn}}
\end{center}
\end{figure}

\begin{figure}[ht]
\begin{center}
\includegraphics[width=3.5 in, angle=-90]{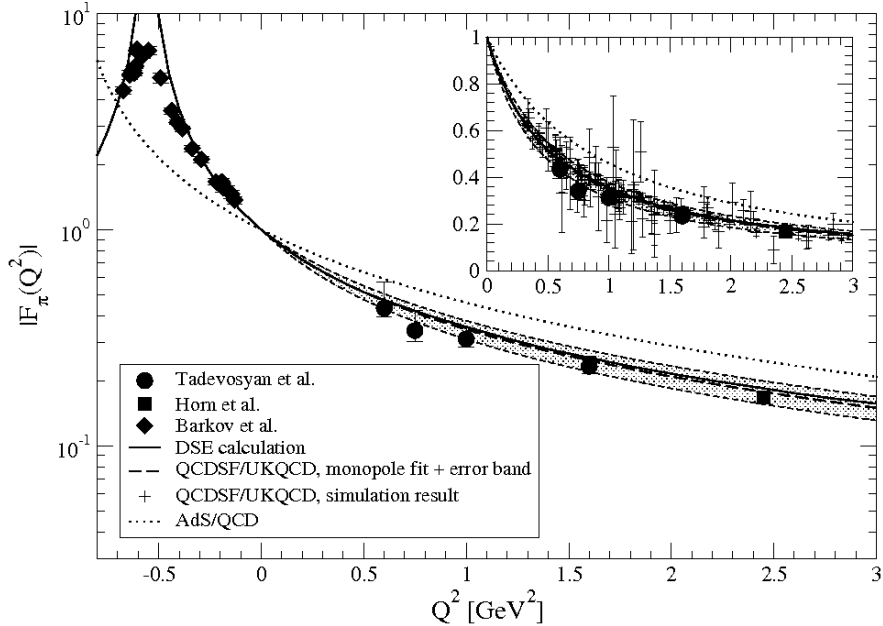}
\caption{Pion form factor data in both the space- and time-like regions.  The three curves represent the Dyson-Schwinger equation calculation (solid), the AdS/QCD calculation (dotted) and the lattice QCD calculations (shaded region).  [Reprinted from \cite{Cloet:2008fw}, with permission from the author.]
\label{fig:piff}}
\end{center}
\end{figure}

The pion elastic form factor is very interesting since non-perturbative calculations can be performed for this relatively simple system.  In addition, the asymptotic limit at infinitely high $Q^2$\footnote{As we focus on space-like momentum transfers, we follow the convention that $-q^2$ = $Q^2$ $>$ 0.} is known \cite{Lepage:1979zb, Farrar:1979aw} and is given by

\begin{equation}
\label{fpi}
F_\pi(Q^2) \stackrel{Q^2\rightarrow \infty} {\longrightarrow} \frac {16\pi\alpha_s(Q^2)f_\pi^2}{Q^2},
\end{equation}
where $\alpha_s$ is the strong coupling constant and $f_\pi$ is the pion decay constant.
The $Q^2$ dependence of this form factor is consistent with the constituent counting rule for electron elastic scattering from the pion.
An interesting way to gauge the transition region between hadronic and partonic degrees of freedom might be from the quark mass itself. Theoretical studies \cite{GutierrezGuerrero:2010md} of the pion form factor indicate that the running of the quark mass is an important ingredient in the calculations. We know that Bjorken scaling \cite{Bjorken:1967, Bjorken:1969ja, Feynman:1969} sets in at relatively low values of momentum transfer in deep inelastic scattering, i.e.\ when more than $\approx$ 2\ GeV/$c$ is imparted to the quark.  From Figure~\ref{fig:dsmass} it is noted that the quark mass is already near its current quark mass at 2\ GeV/$c$.  The pion form factor presents an interesting test case since the pion is only a quark-antiquark system.  To impart an average of 2\ GeV/$c$ to a quark and antiquark, only 16\ GeV$^2$ need be imparted to the pion.  This should be achievable or nearly achievable in both the space-like and time-like regions, defined in Figure~\ref{fig:pifeyn}. Note that time-like momentum transfers, this argument is invalid in regions where high-mass resonances modify the form factor. The space-like data at very high $Q^2$ make use of the process indicated in Figure~\ref{fig:pifeyn}(a), i.e.\ electron scattering from the virtual pion cloud in the proton.  Time-like data are from the process in Figure~\ref{fig:pifeyn}(b), where the $e^+e^- \rightarrow \gamma^* \rightarrow \pi^+\pi^-$ reaction is employed.  Existing precision data in the space-like region \cite{Amendolia:1986wj,Dally:1981ur,Dally:1982zk,Tadevosyan:2007yd,Blok:2008jy,Huber:2008id} and a sample of data in the time-like region \cite{Barkov:1985ac} for the pion elastic form factor are shown in Figure~\ref{fig:piff}.  Three disparate theoretical approaches \cite{Grigoryan:2007wn,Cloet:2008fw,Brommel:2006ww} are also represented in the figure.  The Dyson-Schwinger equation calculations should approach the pQCD limit at very high momentum transfer, while the AdS/QCD approach will give at least the same $Q^2$ dependence. (AdS/QCD attempts to solve the strong coupling theory of QCD with the string-theory inspired technique of instead 
solving a dual theory with weak coupling in 5-dimensional space. See, e.g., \cite{deTeramond:2008ht}.)
For an informative review of the space-like form factor data, see \cite{Huber:2008id} and for an excellent theoretical review, see \cite{Bakulev:2004cu,Cloet:2008fw}.   

At very low values of $Q^2$, the form factor was measured \cite{Amendolia:1986wj,Dally:1981ur,Dally:1982zk} by scattering real pions from electrons in a target.  However, at high values of $Q^2$, the pion space-like form factors are deduced from electron scattering from a virtual pion in a proton target.  Highly precise data \cite{Tadevosyan:2007yd,Blok:2008jy,Huber:2008id,Volmer:2000ek} have been taken only up to a momentum transfer of 2.5\ GeV$^2$ at Jefferson Lab.  When the JLab facility is upgraded to 12 GeV, data up to 6\ GeV$^2$, where the hard and soft processes become comparable, should be possible.  Presently, two high $Q^2$ values for the time-like pion form factor have been reported \cite{Milana:1993wk,Seth:2009dk} at 9.6 and 13.48\ GeV$^2$.  Although one might expect that pQCD would begin to dominate at these values of momentum transfer, the results are $Q^2F_\pi = 0.94 \pm 0.08$ and $1.01\pm0.11\pm 0.07\ GeV^2$, respectively, much larger than the value of $\approx$0.10 GeV$^2$ expected for pQCD as given by \ref{fpi}; the prediction for space-like and time-like form factors should be the similar. The large value of the time-like form factor indicates that the process is primarily non-perturbative or that resonances have a strong influence even at this high value of $q^2$.  The prospect for improving the measurements in the time-like region is excellent because of the $e^+e^-$ colliders in operation or recently in operation.

\subsubsection{Pion transition form factor}

The lowest order diagram that describes the $e^+e^- \rightarrow e^+e^-\pi^\circ$ process is shown in Figure~\ref{fig:pifeyn}c.  
The pion transition form factor in lowest order pQCD can be determined from

\begin{equation}
\label{fpig}
Q^2F_{\pi\gamma}(Q^2) =  \frac{\sqrt{2}f_\pi}{3}\int_0^1\frac {dx}{x}\phi_\pi(x,Q^2),
\end{equation}
where $f_\pi$ is the pion decay constant, $x$ is the momentum fraction for a parton in the pion, and $\phi_\pi$ is the parton distribution amplitude for a parton in the pion.
The pion transition form factor has traditionally been cited as the best example of the approach to a pQCD limit.  The process has an asymptotic limit \cite{Lepage:1980fj} that is much larger than that of the pion form factor:

\begin{equation}
\label{fpqcd}
Q^2F_{\pi\gamma}(Q^2) \stackrel{Q^2 \rightarrow \infty} {\longrightarrow} \sqrt{2}f_\pi.
\end{equation}

Recent results \cite{Aubert:2009mc,Dubrovin:2010zz} from the BaBar Collaboration for the $e^+e^- \rightarrow e^+e^-\pi^\circ$ process have been extended to a $Q^2$ of $\approx$40 GeV$^2$ and surprisingly these results do not exhibit a $Q^{-2}$ dependence for the form factor expected from pQCD. 
Nevertheless, some authors \cite{Dorokhov:2010zz,Dorokhov:2010zz2,Gorchtein:2011vf,Pham:2011jq} have described the data  by using QCD-inspired models.
Recent works \cite{Roberts:2010rn,Chang:2011vu} argue strongly that reasonable nonperturbative descriptions of this process should approach the pQCD limit from below the limit, a perspective also developed elsewhere \cite{Mikhailov:2009sa,Brodsky:2011yv,Bakulev:2011rp,Brodsky:2011xx}. 
These results appear to cast doubt about the data which exceed the limit at such high values of $Q^2$.  Moreover, recent BABAR data \cite{sanchez:2011hk,Lees:2010de} for the transition form factors of the $\eta$, $\eta'$ and $\eta_c$ appear to be described by pQCD treatments at high Q$^2$.
As yet unpublished data from the Belle Collaboration \cite{Uehara:2012ag} are below the BaBar result,
more consistent with the high $Q^2$ asymptotic limit.

\subsection{The Nucleon}

\subsubsection{Elastic Form Factors}
\label{sec:nucleonformfactor}

A recent review of the electromagnetic nucleon
form factors is \cite{Arrington:2006zm}.
Here we focus on the high $Q^2$ behaviour of the form factors.
We consider the helicity conserving Dirac $F_1$ and 
helicity nonconserving Pauli $F_2$ form factors, or equivalently
the electric and magnetic form factors, $G_E = F_1 - \tau F_2$
and $G_M = F_1 + F_2$, respectively, with $\tau = Q^2/4m_p^2$.
Ignoring logarithmic corrections and running of the strong coupling
constant $\alpha_s(Q^2)$,  the constituent counting rules and
perturbative QCD  \cite{Lepage:1980fj}
predict that $F_1$ falls as 1/$Q^4$, and $F_2$ falls as 1/$Q^6$, 
so $G_M$ also falls as 1/$Q^4$.
While the magnitudes of the form factors at $Q^2 \to \infty$ are not 
known, with reasonable assumptions $G_M^n / G_M^p \to -2/3$.
Following our arguments above, one might expect that the proton form factors
become asymptotic for $Q^2$ $\approx$ 36 GeV$^2$.

To date, the ranges of measurements for the various form factors are limited to $Q^2$
$\approx$30 GeV$^2$ for $G_M^p$ \cite{Arnold:1986nq},
5 GeV$^2$ for $G_M^n$  \cite{Lachniet:2008qf},
8.5 GeV$^2$ for $G_E^p$\cite{Puckett:2010ac}, and
3.4 GeV$^2$ for $G_E^n$ \cite{Riordan:2010id},
so one would not expect them to be in the perturbative regime.
However both magnetic form factors follow the dipole formula, 
$G_M^{p,n} = \mu_{p,n} G_D = (1+Q^2/0.71)^{-2}$,
which has the expected high-$Q^2$ scaling, within about 10\%.
Furthermore, with similar precision, at all $Q^2$
$G_M^n/G_M^p$ $\approx$ $\mu_n/\mu_p$ = -0.685, which is 
consistent with the predicted ratio of -2/3.
In contrast, estimates of the actual magnitude of the perturbative QCD 
contribution to the proton magnetic form factor \cite{Isgur:1984jm}
indicate that it is likely small, perhaps 1\% of $G_M^p$.

\begin{figure}[ht]
\begin{center}
\includegraphics[height=2.3 in, angle=0]{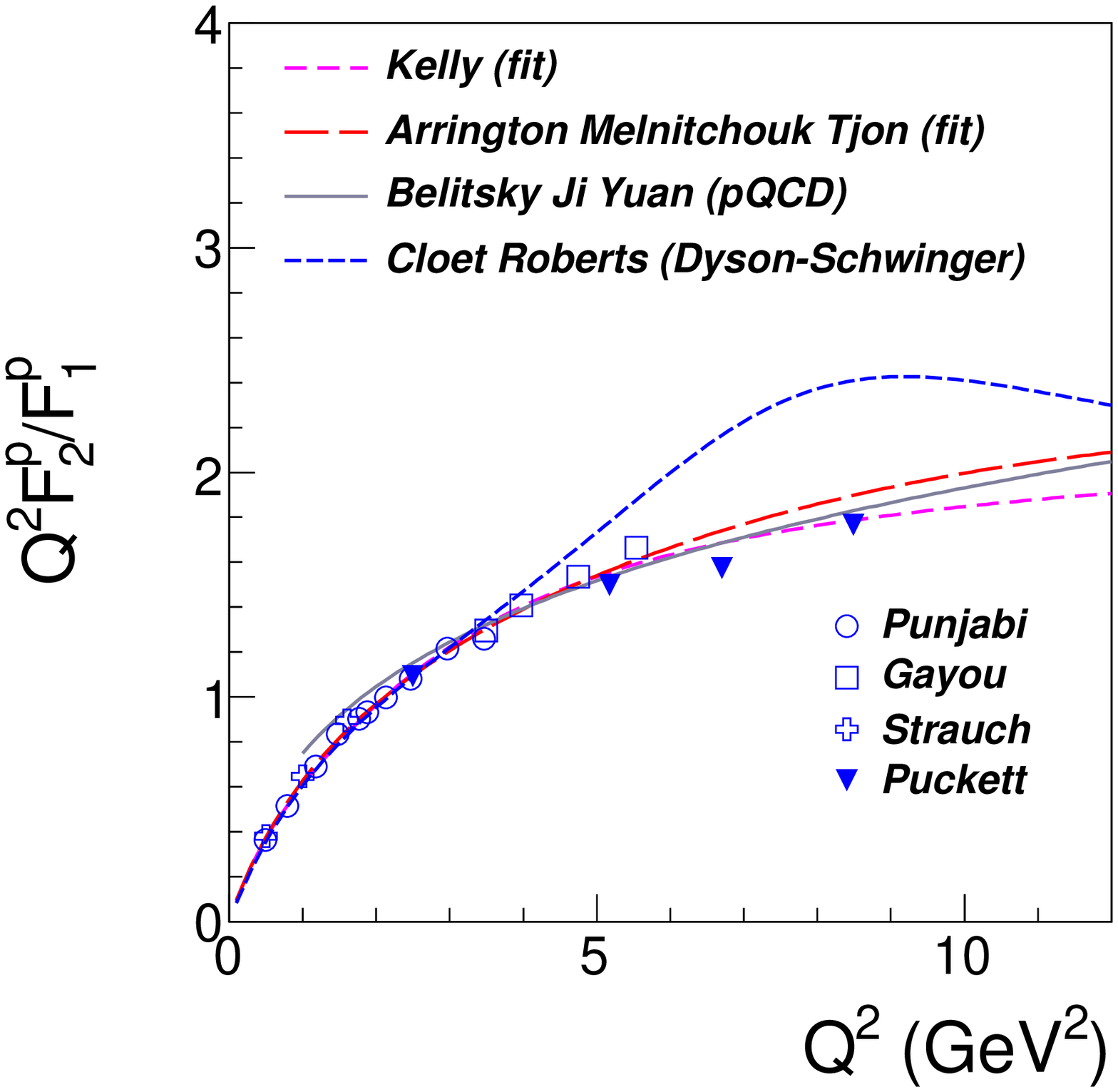}
\includegraphics[height=2.3 in, angle=0]{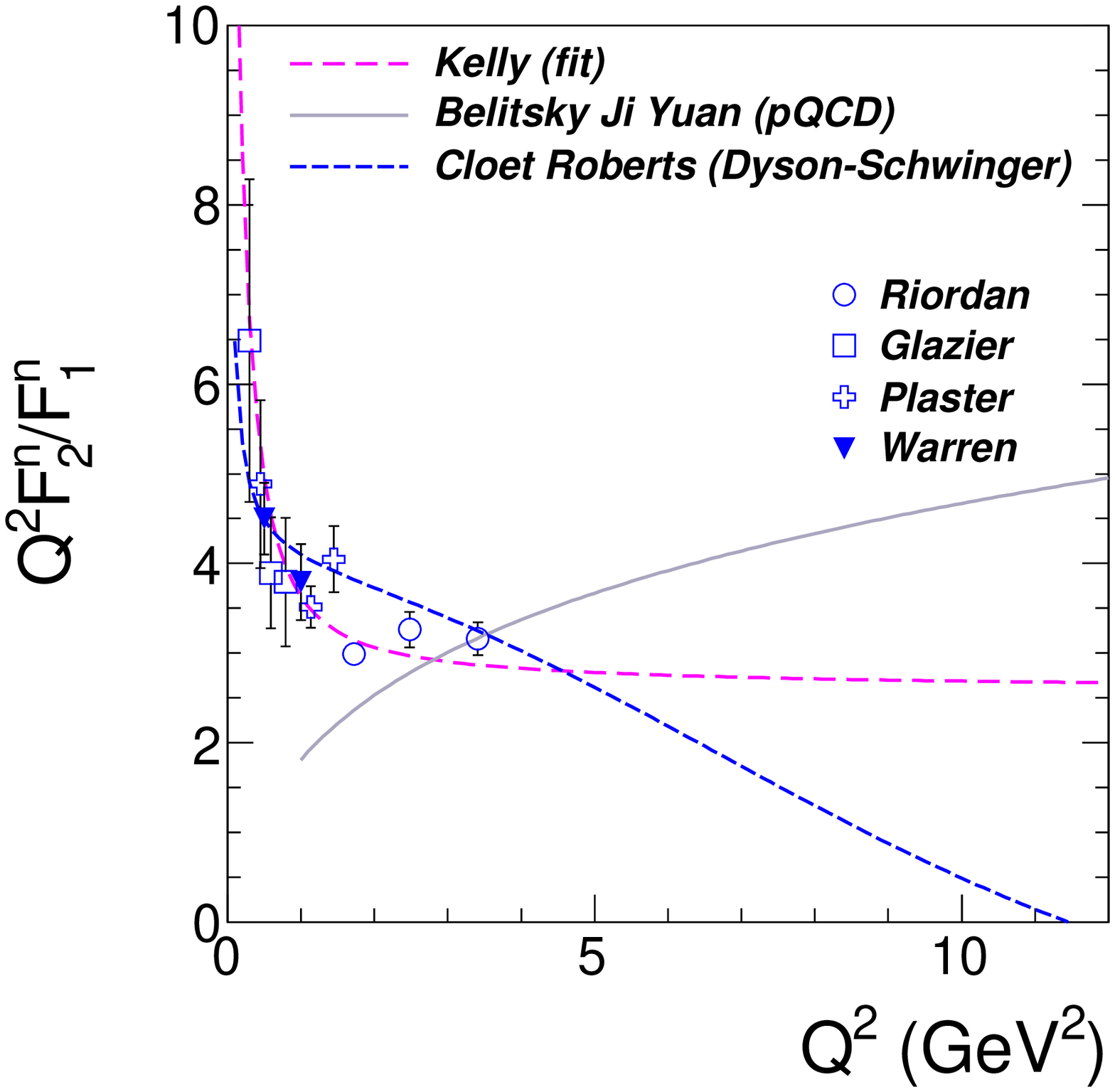}
\caption{The form factor ratio $Q^2 F_2 / F_1$ for a restricted set of
data for the 
proton (left) and the neutron (right) compared to
the Kelly fit \cite{Kelly:2004hm},
the ``AMT'' fit for the proton \cite{Arrington:2007ux}, 
a recent Dyson-Schwinger calculation
\cite{Cloet:2008re,Chang:2010hb,Chang:2011tx}, 
and  the ``BJY'' pQCD scaling function from \cite{Belitsky:2002kj}
with $\Lambda$ = 0.15 GeV, normalized to the higher $Q^2$ data.
Increasing $\Lambda$  to 0.20 GeV would basically overlap the 
AMT fit, while decreasing $\Lambda$  to 0.10 GeV would basically overlap the 
Kelly fit. The ``BJY'' parameterization cannot work for the neutron unless
the ratio increases.
Data are from 
Punjabi \cite{Punjabi:2005wq}, Gayou \cite{Gayou:2001qd}, 
Strauch \cite{Strauch:2002wu}, Puckett \cite{Puckett:2010ac},
Riordan \cite{Riordan:2010id}, Glazier \cite{Glazier:2004ny}, 
Plaster \cite{Plaster:2005cx}, and Warren \cite{Warren:2003ma}.
\label{fig:qsqf2of1}}
\end{center}
\end{figure}

The electric form factors do not follow the dipole formula;
the falloff of $G_E^p(Q^2)/G_D(Q^2)$ is well known -- this
disagrees with the scaling expectation \cite{Brodsky:1974vy}
that $G_E/G_M$ $\to$ constant.
We consider instead the ratio $F_2/F_1$.
Using $R = G_E/G_M$, $F_2/F_1 = (1-R)/\kappa(\tau + R)$.
(We normalize $G_M(0)$ = $\mu$ but $F_2(0)$ = 1.)
In pQCD, neglecting orbital angular momentum contributions, helicity flip 
costs a power of $Q^2$ so that one expects $Q^2 F_2 / F_1 \, \to \, {\rm constant}$.
But since the first JLab $G_E^p$ data appeared \cite{Jones:1999rz} it has been known that this
formula does not work well in the range of measured data; 
instead $Q F_2 / F_1 \, \approx \, {\rm constant}$.
This result was explained with quark models as indicating the
importance of relativity and orbital angular momentum of the 
quarks in the proton \cite{Miller:2002qb}.
A refined pQCD analysis including orbital angular momentum suggests a modified scaling,
$Q^2 F_2 / F_1 \, \propto \, \ln^2(Q^2/\Lambda^2)$, with $\Lambda$ a constant
\cite{Belitsky:2002kj}. (See also \cite{PhysRevD.69.076001}.)
Figure~\ref{fig:qsqf2of1} shows that this formula works quite well for
the proton, but it does not work at all for the neutron.
The Dyson-Schwinger calculation, drawn from
\cite{Cloet:2008re,Chang:2010hb,Chang:2011tx},
has been extended up to 12 GeV$^2$ for the first time as shown in
Figure~\ref{fig:qsqf2of1}.  
The agreement up to 5 GeV$^2$ is quite good, but the deviation from 
the data for the proton is dramatic. 
A possible refinement to this calculation is to choose a quark mass 
function (See Figure~\ref{fig:dsmass}.) that has a different falloff
rate for the quark momentum; the ratio might be a sensitive probe
of the momentum dependence of the dressed quark mass function.

Of equal importance to the space-like form factors measured with
electron scattering are the time-like form factors measured in
colliders through reactions such as $p \overline{p} \to e^+ e^-$. 
The cross section is given by
\begin{equation}
{{d\sigma}\over{d\Omega}} = {{\alpha^2}\over{2s\sqrt{1-4m_p^2/s}}}
\left[(1+\cos^2\theta) G_M^2(s) + {{4 m_p^2}\over{s}} \sin^2\theta G_E^2(s) \right],
\end{equation}
where $\theta$ is the outgoing electron angle,
and Mandelstam $s = q^2 = -Q^2$
is the photon virtuality.

While it might appear that the differing angle dependences of the
electric and magnetic terms make separations easy, the low luminosity
of experiments coupled with small cross sections and large backgrounds
has in general prevented separations of $G_E$ and $G_M$. 
Instead it is typically assumed either that $G_E = 0$ or $G_E = G_M$.
The estimated timelike proton magnetic form factor for 
$q^2$ $>$ 8 GeV$^2$ appears to roughly scale as
expected from pQCD, with $q^4 G_M$ $\propto$ $\alpha_{strong}^2$.
However, from pQCD it is expected that $G_{M\, timelike}(q^2) = G_{M\,
  spacelike}(Q^2)$, while experimentally the 
timelike form factor is about a factor of two larger.

To summarize, even though existing data are not expected to be 
in the perturbative regime, the magnetic form factors agree 
reasonably well with the expected pQCD scaling. 
The proton form factor ratio can be considered to be in 
agreement as well, if orbital angular momentum is included.
The neutron electric form factor does not agree with
perturbative expectations, nor does the ratio of timelike to
spacelike form factors.
Since the form factor magnitudes appear to be largely nonperturbative,
the agreements in the scaling behaviour might be fortuitous.
While it is beyond our scope to address in any detail, 
the form factor data can be qualitatively understood 
through various quark models or parameterized GPDs.

\subsubsection{Hard Compton scattering at high energy}

\begin{figure}[ht]
\begin{center}
\includegraphics[height=1.75 in, angle=0]{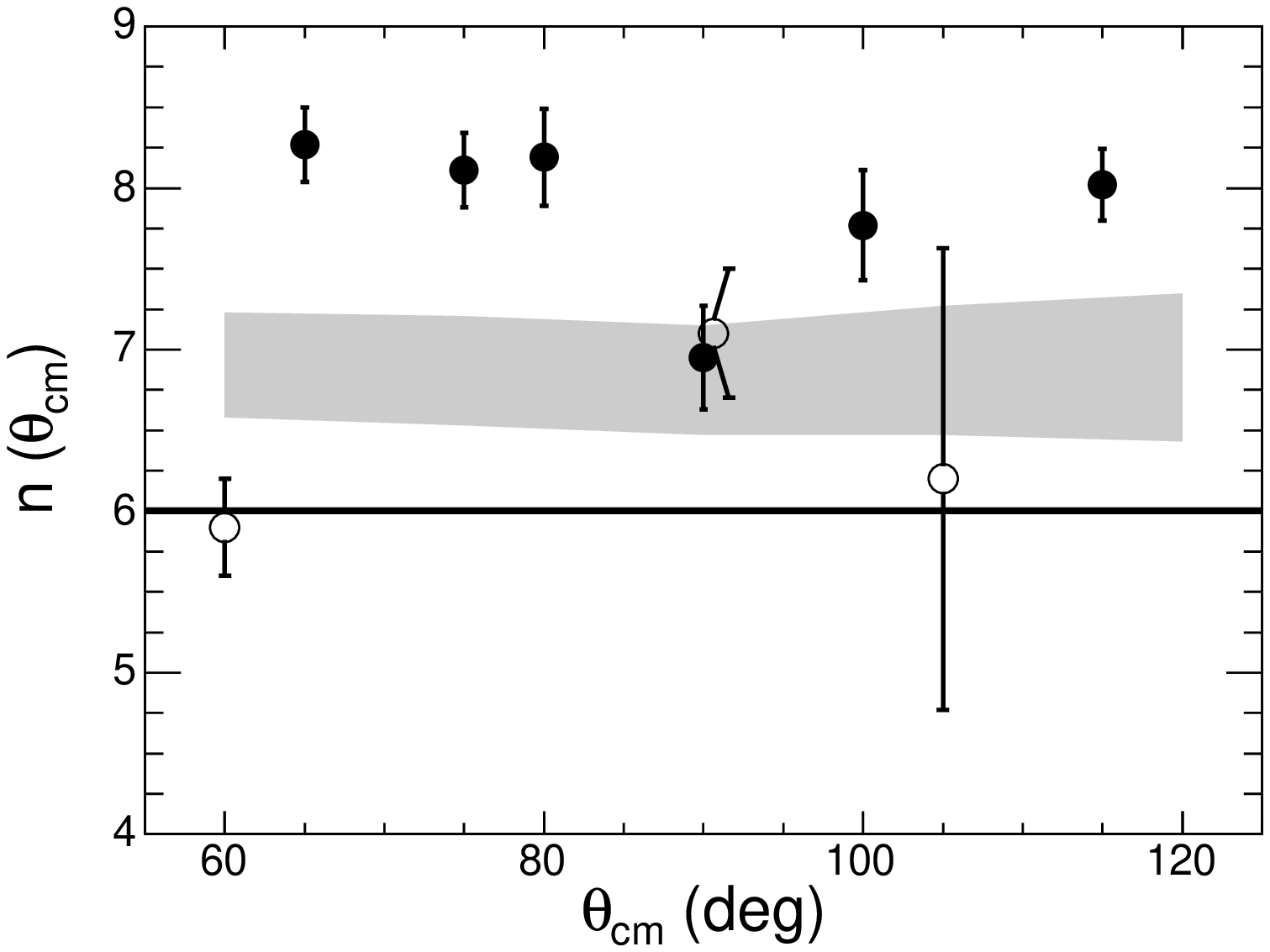}
\includegraphics[height=1.75 in, angle=0]{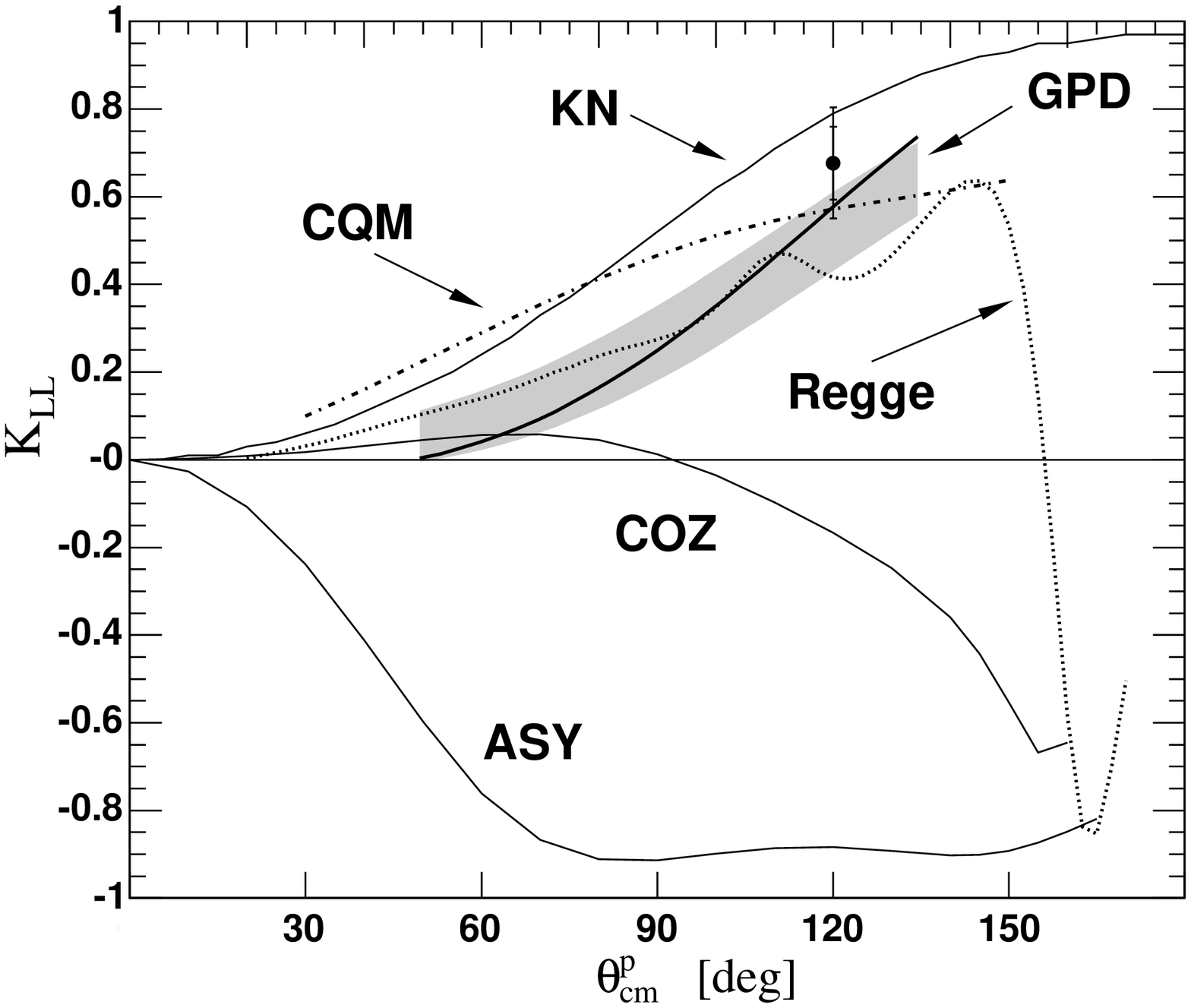}
\caption{Left: Scaling exponent $n$ in $d\sigma/dt(\theta_{cm}) \propto s^{-n}$ 
as a function of $\theta_{cm}$ for real Compton scattering.
Open circles are from \cite{Shupe:1979vg},
closed circles are from \cite{Danagoulian:2007gs},
and the gray band represents the scaling exponent predicted by GPD models.
The line at $n$ = 6 is the expectation from pQCD.
Taken from \cite{Danagoulian:2007gs}.
Right: Longitudinal polarization transfer coefficient $K_{LL}$,
compared to several calculations.
The curve labelled ``KN'' is the Klein-Nishina result, for polarization
transfer to a point-like spin-1/2 particle.
The curves labelled ``ASY'' and ``COZ'' are pQCD calculations with
two different choices for the distribution amplitude.
Other calculations include constituent quark (``CQM''), generalized
parton distribution (``GPD''), and Regge (``Regge'') models.
[Reprinted Figure~4 with permission from \cite{Hamilton:2004fq}. Copyright
(2005) by the American Physical Society.]

\label{fig:rcs}}
\end{center}
\end{figure}

The real Compton scattering (RCS) reaction is $\gamma p \to \gamma p$.
The perturbative QCD prediction for hard (Mandelstam $s$, $-t$, and $-u$ $\gg$ $m_p^2$) 
RCS is $d\sigma/dt(\theta_{cm}) \propto s^{-6}$.
As shown in Figure~\ref{fig:rcs},
this prediction was roughly supported by cross section data from Cornell
\cite{Shupe:1979vg} for $E_{\gamma}$ = 2 -- 6 GeV, but a subsequent more
comprehensive Jefferson Lab experiment 
\cite{Danagoulian:2007gs} found the scaling is more consistent with $s^{-8}$;
oddly both experiments find $s^{-7}$ scaling at $\theta_{cm}$ = 90$^{\circ}$.
Differences between the two results could be explained if there was an
energy-dependent leakage of $\gamma p \to p \pi^0$ events into the RCS 
channel in the Cornell data, as $\pi^0$ production is about two orders of 
magnitude larger at these energies.

\begin{figure}[ht]
\begin{center}
\includegraphics[height=1.0 in, angle=0]{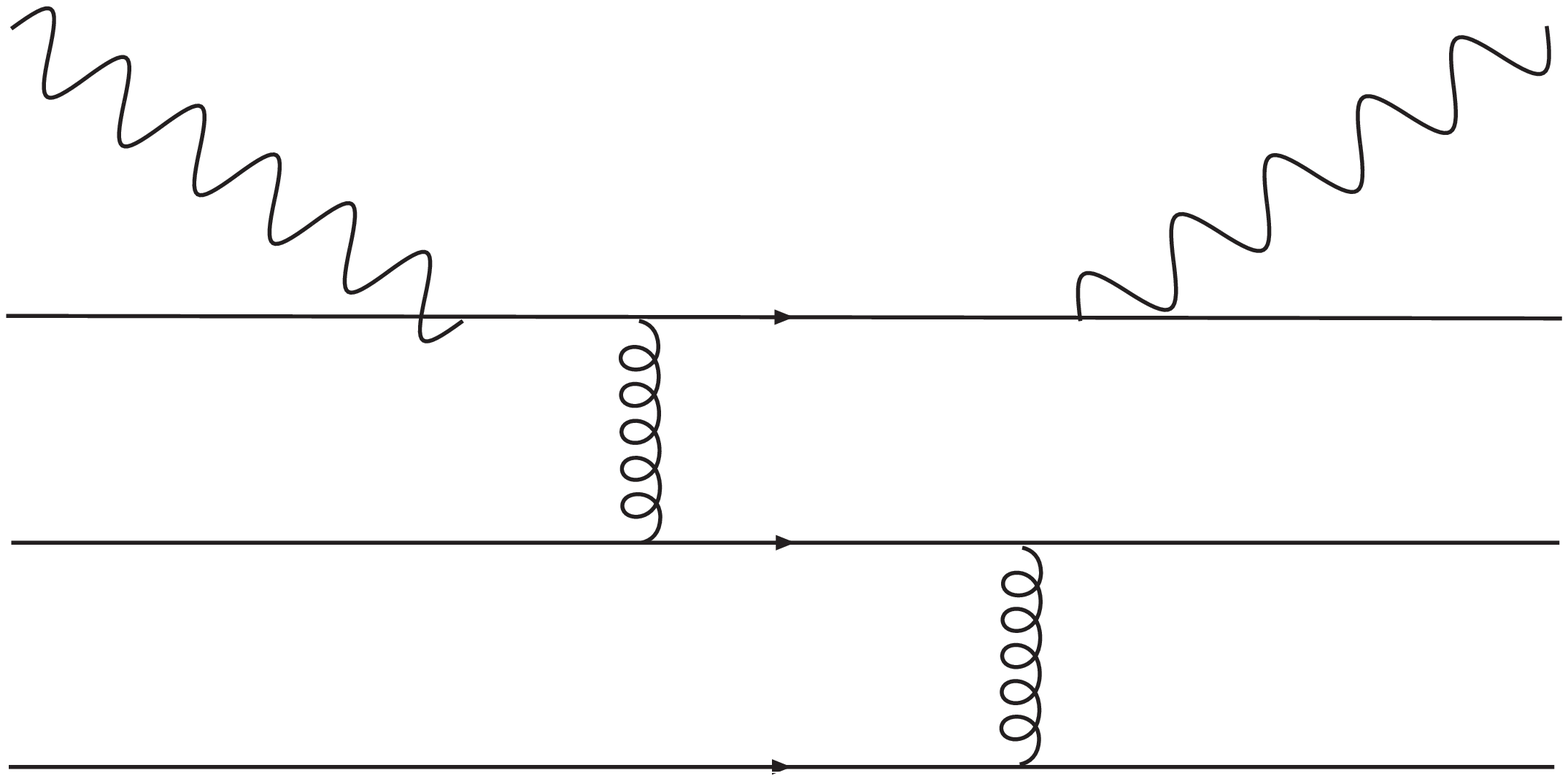}
\hspace*{1cm}
\includegraphics[height=1.0 in, angle=0]{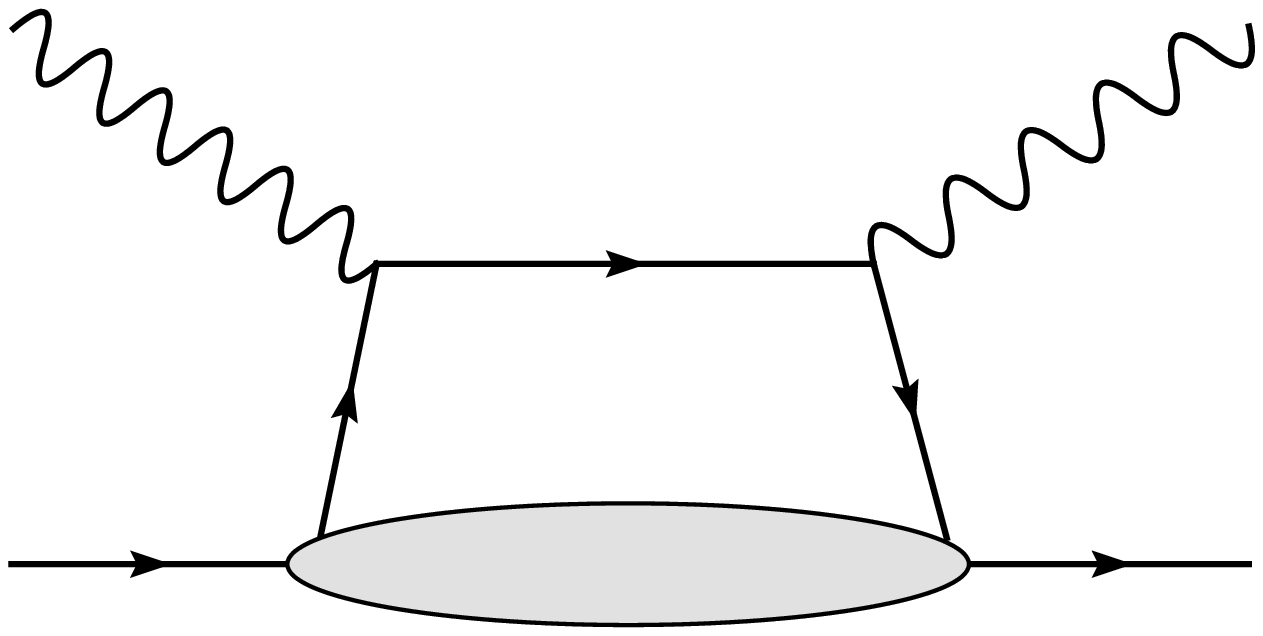}
\caption{Left: A sample pQCD Feynman diagram for Compton
scattering, with the minimal two hard gluon exchanges to share the
absorbed photon momentum among the quarks.
The absorbed and emitted photons can couple to different quarks.
Right: The handbag diagram for RCS, where the absorbed and emitted
photon attach to the same quark line, and the momentum is shared
with other constituents of the nucleon through the wave function
(soft gluon exchanges).
\label{fig:rcsdiag}}
\end{center}
\end{figure}

The most recent pQCD calculation of RCS is given in \cite{Thomson:2006ny},
which reviews and compares to earlier work.
A sample Feynman diagram is shown in Figure~\ref{fig:rcsdiag}.
If the RCS calculation is normalized using the ratio to the proton form factor,
then the RCS calculations are only a factor of several below the data;
the factor decreases with energy due to the faster energy dependence of the
data. Polarization transfer coefficients were measured in
\cite{Hamilton:2004fq}, see  Figure~\ref{fig:rcs}, but for $E_{\gamma}$ = 3 GeV and  
$\theta_{cm}$ = 120$^{\circ}$, corresponding to Mandelstam $-u$ = 1.1 GeV$^2$,
which is too small to expect pQCD to apply.

There have been several attempts to describe RCS through the handbag
mechanism, shown in Figure~\ref{fig:rcsdiag},
such as in a constituent quark model \cite{Miller:2004rc}
and with GPDs \cite{Huang:2001ej,Kroll:2007fu}, 
in which RCS depends on $1/x$ moments of the
GPDs.
While the validity of factorization in the 
GPD approach to real photon reactions has been questioned, it is
addressed in \cite{Huang:2001ej,Kroll:2007fu}.
Generally, there has been sufficient flexibility in these approaches to at least qualitatively,
but consistently explain the nucleon form factor and RCS data.
There appears to be no simple explanation of why the scaling has 
$n$ = 8\footnote{Prior to the appearance of the JLab data,
it was noted \cite{Huang:2000kd} that in the vector meson dominance picture
the photon couples through its hadronic component -- e.g. 
the $\rho$ meson -- which would naturally
lead to an $s^{-8}$ energy dependence.
But it was argued that the Cornell data were consistent with $s^{-6}$,
and the VMD contribution was estimated to be perhaps 10\% the size of
the data.}.
These approaches also explain the polarization transfer measurement.
While these model calculations for the polarization transfer tend to qualitatively
resemble the Klein-Nishina result, as shown in Figure~\ref{fig:rcs}, apparently
the interferences between various diagrams in the pQCD calculations lead
to the full calculation being very roughly opposite in sign to the
Klein-Nishina formula.

In summary, it appears that RCS cannot be explained purely
perturbatively.
It might be explained with the perturbative
scattering of a photon and quark, with soft nucleon-structure physics
modeled through either quark models or GPDs, but more work needs
to be done on improving the energy dependence.

\subsubsection{Deeply Virtual Compton Scattering}

\begin{figure}[ht]
\begin{center}
\includegraphics[height=1.25 in, angle=0]{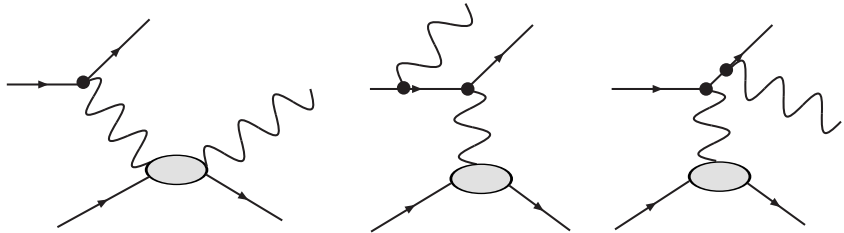}
\caption{Left: Virtual Compton scattering.
Middle: The Bethe-Heitler process, with a photon emitted
by the incoming electron. 
Right: The Bethe-Heitler process, with a photon emitted
by the outgoing electron.
\label{fig:vcsbh}}
\end{center}
\end{figure}

Virtual Compton Scattering (VCS) is a generalization of RCS, in which
a virtual photon emitted by a scattered electron is absorbed by a nucleon,
with a real photon emitted -- see Figure~\ref{fig:vcsbh}.
Deep VCS (DVCS) refers to this process at high $Q^2$.
The competing Bethe-Heitler (BH) process, in which electrons passing
near the nucleus radiate photons, is understood and calculable.
Rather than being an annoying background, the BH process is an advantage;
similar to the case of holography it can be thought of as providing 
a reference beam that gives us additional information.
The interference of DVCS and BH allows the phase of the DVCS
amplitude to be determined.
Note that the BH photons are emitted generally in the direction
of the emitting electron, and the DVCS process becomes increasingly
dominant with increased energy.

Interest in the DVCS process burgeoned with the realization
that it could provide important information on GPDs and the
total angular momentum of quarks in the nucleon \cite{Ji:1996ek}
for large $Q^2$ and small $-t$. 
The pQCD diagram and the handbag diagram, which is assumed
in the GPD approach, are the same as in Figure~\ref{fig:rcsdiag}, 
except that the incoming photon is virtual.
While the overwhelming majority of calculations have used the
GPD framework, the validity of this approach can be studied to
some degree with pQCD calculations.
In \cite{Thomson:2006ny}, the DVCS process was calculated and
the approximation that the incoming and outgoing photon interact
with the same quark was studied. 
The two photons attaching to the same quark line was dominant
for photon scattering angles up to 20$^{\circ}$.

A recent review that discusses DVCS data and GPDs is
given in \cite{Hyde:2011ke}. 
The VCS amplitude in leading order depends on integrals 
of the GPDs $H$, $\tilde{H}$, $E$, and $\tilde{E}$ weighted by 
kinematic factors.
The major observables studied have been cross sections,
beam-helicity dependent cross sections or asymmetries, 
and longitudinally polarized target asymmetries.
The data are generally in the range $Q^2$ $\approx$ 1 -- 3 GeV$^2$
and $-t$ $<$ 1 GeV$^2$ -- note that the four-momentum transfer $-t$
is not the same as the photon four-momentum $Q^2$ as the final
state includes $p + \gamma$.
Both the neutron and the proton have been studied.
At present, the various measurements tend to be qualitatively
consistent with GPD models that include some amount of 
higher order twist-3 contribution, but there is 
no comprehensive, quantitative explanation.

\subsubsection{Photo-pion reactions}

Meson production reactions were among the first pieces of evidence
for the constituent count rules \cite{Anderson:1976ph,Clifft1977144}, yet these
reactions have been notoriously difficult to calculate.
As pointed out in the pioneering work of \cite{Farrar:1990eg}, 
pion photoproduction calculations require several thousand 
Feynman diagrams.
The calculated cross section 
has a large sensitivity to the baryon wave functions used, 
is similar in size to the experimental data, and 
has a potentially interesting helicity structure.
However, the numerical techniques used were related to those
used in a calculation of Compton scattering 
\cite{Farrar:1989vp} which is not in agreement 
with subsequent work -- see \cite{Thomson:2006ny}.

The pQCD calculation is simplified to only hundreds of diagrams
in the quark-diquark  model of the nucleon \cite{Kroll:1996vz,Folberth:1996vv}.
Diquarks may be viewed as an effective quasi-elementary particles that
incorporate some nonperturbative physics of the nucleon, for reactions
in which the interaction is primarily with a single quark.
There are spin-0 scalar diquarks and spin-1 diquarks.
Photoproduction of $K^+\Lambda^0$ is most studied as  it involves
only scalar diquarks -- the spin of the $\Lambda^0$ is usually viewed
as being carried by the $s$ quark.
The prediction of \cite{Folberth:1996vv} for the 
$\gamma p \to \pi^+n$ reaction are roughly of similar size to the data.
A quantitative explanation would require, e.g., additional $u$-channel
processes in the case of the asymptotic distribution amplitude.

More recently there have been GPD based calculations of meson
photoproduction \cite{Huang:2000kd}.
The calculations are about two orders of magnitude below 
the data.
The authors argue that in the GPD picture the formation of a 
meson likely reduces the cross section compared to the emission 
of a photon as in RCS. Since however the data show meson production
cross sections are much larger, they suggest other physics must be 
responsible.
The approximate validity of VMD relations between $\rho$
photoproduction and $\pi p$ scattering, the possible
$s^{-8}$ scaling of $\pi^0$ photoproduction, and the large cross
sections suggest that the VMD picture explains meson photoproduction
for several GeV incident photons.

\begin{figure}[ht]
\begin{center}
\includegraphics[height=2.5 in, angle=0]{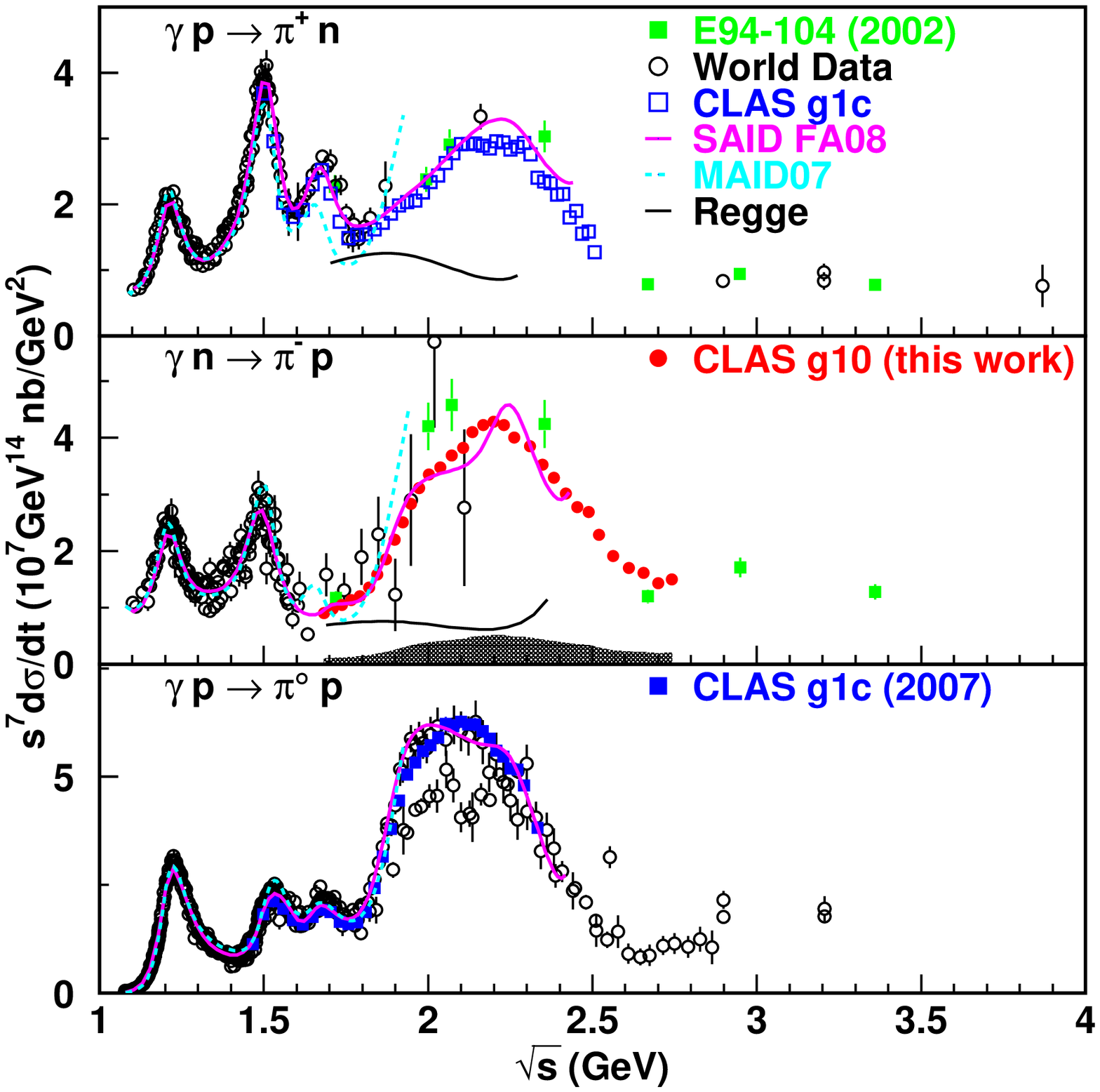}
\includegraphics[height=2.25 in, angle=0]{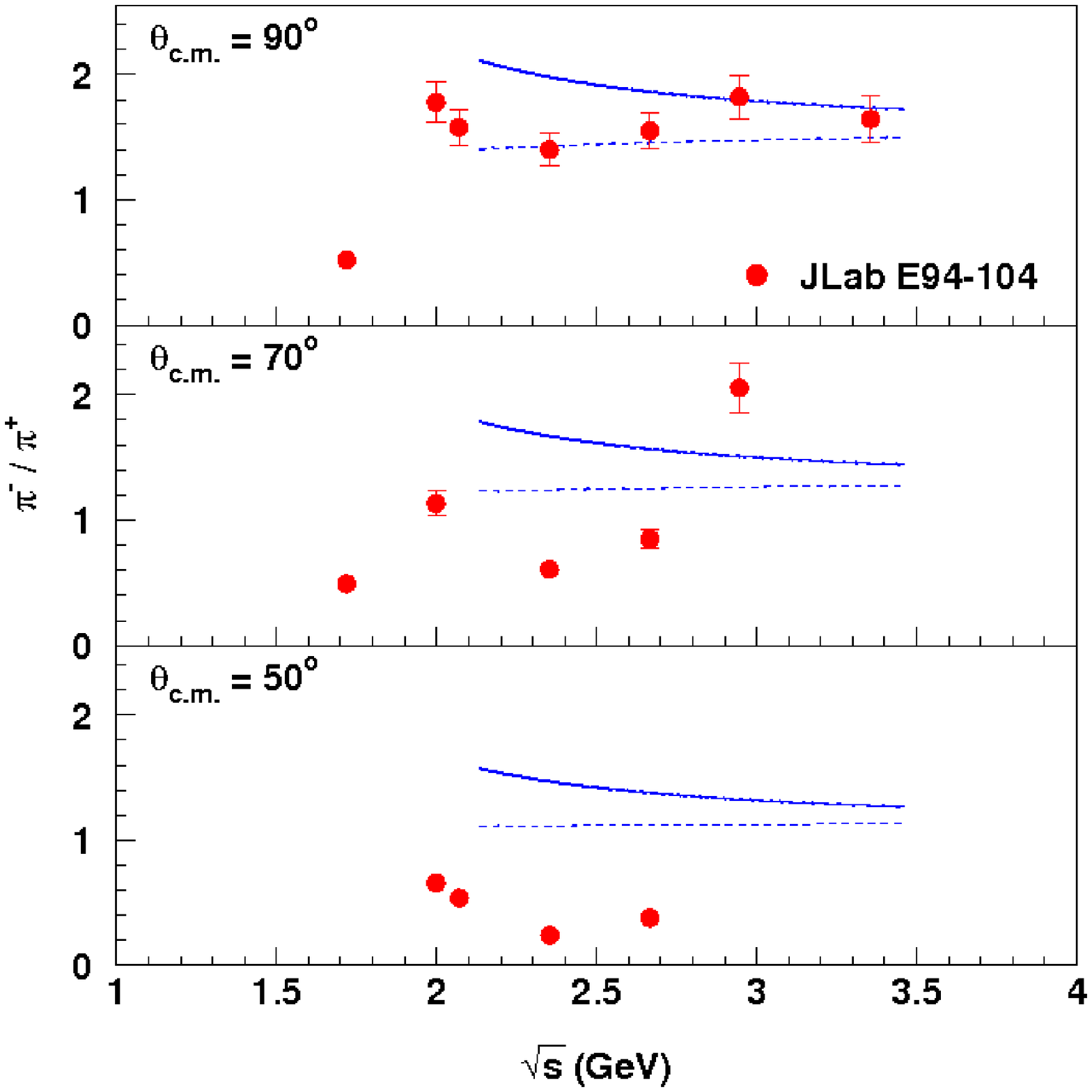}
\caption{
Left: Cross sections for single pion photoproduction at
$\theta_{cm}$ = 90$^{\circ}$
including recent CLAS data
[Reprinted Figure~2 with permission from \cite{Chen:2009sda}. Copyright
(2009) by the American Physical Society.]
Right: Cross section ratio $d\sigma(\gamma n \to \pi^- p) /
d\sigma(\gamma p \to \pi^+ n)$/
[Reprinted Figure~19 with permission from \cite{Zhu:2004dy}. Copyright
(2005) by the American Physical Society.]
The solid line is from \ref{eq:picsrat}, while the dashed line
incorporates mass corrections by reducing $s$ and $u$ by $m_p^2$.
\label{fig:pipgp}}
\end{center}
\end{figure}

In parallel with these theoretical developments 
Jefferson Lab experiments have improved our knowledge
of meson photoproduction.
In \cite{Wijesooriya:2002uc}, recoil proton polarization was measured
in the $\gamma p \to \pi^0 p$ reaction for photon energies up to
4 GeV (center of mass total energy $W$ $\approx$ 2.9 GeV).
The polarizations were found to vary with energy and angle, and did
not appear to approach any smooth behaviour as expected from quark 
models.
A wide range of single pion photoproduction measurements
have also now been done by the CLAS collaboration
\cite{Chen:2009sda,Dugger:2007bt,Dugger:2009pn},
with a fraction of the data shown in Figure~\ref{fig:pipgp}.
It appears that the resonance region extends up to, and the
scaling region starts at, $W$ = $\sqrt{s}$ $\approx$ 2.6 GeV, much
higher than the conventional $W$ = 2 GeV limit to the 
resonance region.
In \cite{Zhu:2004dy,Zhu:2002su}, the cross sections for 
$\gamma p \to \pi^+ n$ and $\gamma n \to \pi^- p$ were measured
to higher $W$.
Figure~\ref{fig:pipgp} shows that
the ratio of the two processes at the highest energies, 
but only at $\theta_{cm}$ = 90$^{\circ}$,
agrees with simple quark estimates \cite{Huang:2000kd,Afanasev:1996mj}:
\begin{equation}
\label{eq:picsrat}
{ {d\sigma(\gamma n \to \pi^- p)} \over {d\sigma(\gamma p \to \pi^+ n)}
}
\approx
\left( {u e_d + s e_u} \over {u e_u + s e_d} \right)^2,
\end{equation}
where $s$ and $u$ are Mandelstan variables and
$e_{u, \, d}$ are the $u$, $d$ quark charges.
Also, the highest energy points in the scaling region appear
to have some oscillation about smooth scaling, perhaps of similar
origin to the behaviour seen in $pp \to pp$ \cite{Dutta:2004fw}.
Thus, there appear to be competing underlying dynamical mechanisms
for the pion photoproduction reactions.

\subsubsection{Baryon transition form factors}

Extracting baryon transition form factors and their asymptotic
behaviour is difficult, as baryon resonances
overlap, are wide, and sit on top of a nonresonant background.
Reliable extraction is aided by polarization measurements,
by high statistics,
by studying multiple decay channels,
and by a dynamical model to get at the bare resonance parameters
from the observed data, as the final state hadrons interact.
There has been an extensive program at JLab aimed at determining 
baryon resonance properties -- see e.g. \cite{Aznauryan:2011ub}.
The most studied case, to the highest $Q^2$, is 
the $N \to \Delta$ transition; it is the only case we consider here.

The $\Delta(1232)$ resonance, probed at low energies, has long been
known to arise from the $L=1$, $J=3/2$, $T=3/2$, or $p_{33}$,
partial wave in $\pi N$ scattering. 
In the constituent quark model the nucleon is photo-excited 
into the $\Delta$ resonance primarily by a quark spin flip; 
with $\Delta J$ = 1, $\Delta L$ = 0, $\Delta S$ = 1, this is 
an $M1$ magnetic dipole transition. 
There is also a small, few percent, electric quadrupole, or $E2$, component. 

The pQCD result that the proton helicity nonflip Dirac and
helicity flip Pauli form factors fall as $Q^{-4}$ and $Q^{-6}$,
respectively, applies to baryon transition form factors,
measured with electroproduction, as well 
\cite{Lepage:1979za}. 
There are several different common conventions for the
three $N \to \Delta$ transition form factors; here we use 
the magnetic dipole $M_{1+}$, electric quadrupole $E_{1+}$ 
and scalar dipole $S_{1+}$. 
The asymptotic expectations for these form factors
are $R_{EM} \equiv E_{1+}/M_{1+}$ $\to$ 1,
and  $R_{SM} \equiv S_{1+}/M_{1+}$ $\to$ constant, as discussed
in \cite{RevModPhys.80.731}, with all falling as $1/Q^4$.
There is no support for an approach to these limits from
data in the measured range of $Q^2$ = 0 $\to$ 8 GeV$^2$.
The magnetic form factor falls faster than the dipole, which
is probably not surprising given the photo-excitation result 
that $R_{EM}$ is small -- there must be a large nonperturbative
component to the spin-flip $M_{1+}$ transition.
But, in addition, $R_{EM}$ $\approx$ 2 -- 3 \% at all $Q^2$; the ratio
gives no clear indication of increasing towards unity.
Finally, $R_{SM}$ gradually drops from about -5\% near the real photon point 
to about -25\% -- see \cite{Aznauryan:2011ub,Gilman:2011zz}.
Thus, there is no indication of an approach toward the
asymptotic predictions.

Nevertheless, several QCD-inspired theoretical approaches -- unitary
transformation \cite{Sato:2009de}, AdS/QCD \cite{Grigoryan:2009pp},
and QCD sum rules \cite{Carlson:1988gt} -- have been applied reasonably 
successfully to this transition given the approximations.  The unitary
transformation approach illustrates the importance of the pion cloud
at low $Q^2$ and the bare nucleon at high $Q^2$, while the QCD sum
rule approach indicates the important cancellations that arise from
the valence quark symmetries of the $N$ and the $\Delta$.


\section {Color Transparency}

For decades, it has been speculated that colour transparency (CT) will emerge from QCD.  In brief, CT occurs when the initial and final state interactions become greatly diminished or vanish in hadron-hadron interactions.  It is widely believed that three conditions must be met for colour transparency to be observed:

\begin{itemize}
\item{A hadron must have been formed in a small size state or point-like configuration (PLC).}
\item{Small size hadrons have small cross sections.}
\item{The small size hadron remains small in size for a significant time during its travel through the nuclear medium.}
\end{itemize}

In particular, searches for the CT effect have been performed for A(p,2p), A(e,e'p), A(e,e'$\pi$) and A(e,e'$\rho$) reactions as well as pion and J/$\psi$ photoproduction reactions and coherent pion-induced jet production on a nucleus.
Thus far, in the A(p,2p) and A(e,e'p) reactions the evidence \cite{Mardor:1998zf,Aclander:2004zm,Makins:1994mm,Garrow:2001di,Rohe:2005vc} for CT has not been convincing.  There are two possible reasons for this: (i) It is inherently difficult to find or produce a nucleon, a three-quark system, in a small state.  A small state for the proton would only occur at extremely high energies where exclusive reactions have little cross section. (ii) At the relatively low energies of these experiments the expansion of the PLC, if it is indeed produced in the first place, occurs within the nuclear medium.

By contrast, the meson being only an antiquark-quark system offers the possibility that the PLC would be more readily formed than that for a baryon \cite{Heiselberg:1991is,Baym:1996kd}.  In fact, evidence has been reported for photoproduction \cite{Dutta:2003mk,Sokoloff:1986bu} of the pion and J/$\psi$ as well as for electroproduction \cite{Airapetian:2002eh,clasie:2007gqa,qian:2009ub,Adams:1994bw,ElFassi:2012nr} of the pion and $\rho$ meson.

Perhaps the most striking evidence for the CT can be found in coherent nuclear processes where a pion diffracts into two jets of high relative transverse momentum \cite{Aitala:2000hc}. The experiment was conducted at FNAL where a 500 GeV pion beam was scattered coherently from targets of C and Pt.  
The results were consistent with the per-nucleus cross section being $\sigma = \sigma_\circ A^{\alpha}$.  A value of $\alpha = 1.6$ was found which is consistent with predictions \cite{Bertsch:1981py,Frankfurt:1993it} of colour transparency. For diffractive processes on single nucleons in the nucleus, the coherent cross section would grow as $A^2$, while the elastic form factor would contribute a factor of $A^{-2/3}$. This would lead to an overall prediction of $\alpha = 4/3$. For normal pion inelastic scattering, one should expect $\alpha = 2/3$.  Thus, the predicted yield ratio between Pt and C is about an order of magnitude more than expected from ordinary diffraction.  This is indeed a strong signal for CT.

A signal for CT is particularly important in indicating when factorization \cite{Brodsky:1994kf,Collins:1996fb,Strikman:2000} occurs in semi-exclusive electro-meson production.  In particular, CT in pion electroproduction is a necessary condition for factorization in exclusive electroproduction of pions.  Exclusive electroproduction of mesons is believed to be an essential tool to access generalized parton distributions at JLab and CERN (COMPASS-II).  The most recent search for the onset of CT in meson electroproduction was performed at JLab.  In this case, a rho was electroproduced in Fe nuclei.  Evidence for a CT effect would be a rise in the transparency of the rho-meson as a function of $Q^2$.  Indeed, evidence for the onset of CT was observed at JLab for the enhancement of transparency of rho mesons at large values of $Q^2$ \cite{ElFassi:2012nr}.
Plans for future experiments at an upgraded JLab are hoped to provide more compelling evidence for the effect in meson electroproduction.  It has been suggested \cite{Miller:2010eh} that COMPASS-II could also provide information on this effect.

\section{The deuteron}

\subsection{Hadronic descriptions of the deuteron}
\label{sec:deuterontheory}

\begin{figure}[ht]
\begin{center}
\includegraphics[width=2.0 in, angle=-90]{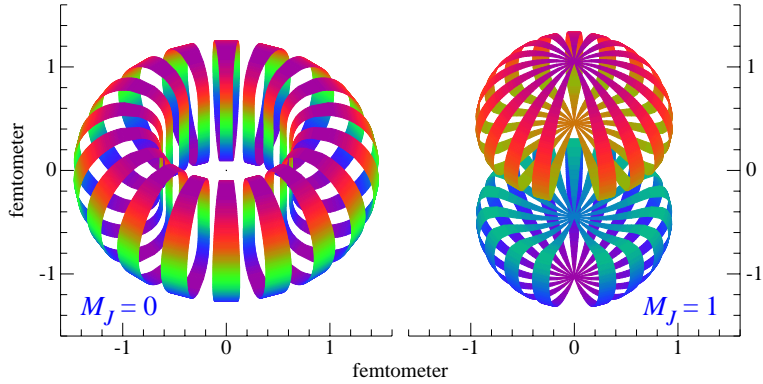}
\caption{Results of a calculation of the surface at a density of 0.24 nucleons/fm$^3$ in the deuteron. Adapted from \cite{Forest:1996kp}.
\label{fig:donut}}
\end{center}
\end{figure}

Gilman and Gross put forward an excellent review \cite{Gilman:2001yh} of the theoretical and experimental status of studies of the deuteron prior to 2001.  Here we present a brief summary of the hadronic description of the deuteron found in this work as well as approaches since 2000. Further, because of the emphasis on the high momentum transfer in this report, we focus primarily on relativistic calculations of electron-deuteron elastic scattering and two-nucleon photodisintegration of the deuteron after 2000. The deuteron is particularly notable for revealing the role of the tensor force in the nucleon-nucleon interaction.  Indeed, the deuteron would not be bound without the tensor force. A particularly good discussion of electron-deuteron elastic scattering that emphasizes the geometric implications of the tensor force on the deuteron structure is given in \cite{Forest:1996kp}.  In particular, if a deuteron can be aligned in a fashion that it is in an $M_J = 0$ magnetic substate, where $J$ is the spin of the deuteron, then the deuteron will have a toroidal shape.  Whereas, if the deuteron is in an $M_J = 1$ or $M_J = -1$ substate, then it will have a ``dumbbell'' shape as shown in Figure~\ref{fig:donut}. The hole in the torus is a reflection of the repulsive core of the N-N interaction, while the overall shapes are largely governed by the relatively strong tensor force below 2 fm. If the deuterons are aligned in these magnetic substates, then these shapes strongly influence electron scattering.  In this way, electron scattering from aligned deuterons is sensitive to the underlying model of the deuteron, in particular, the influence of the tensor force which gives rise to the deuteron $d$- state and leads to the non-spherical shapes.

\begin{figure}[ht]
\begin{center}
\includegraphics[height=2.5 in, angle=0]{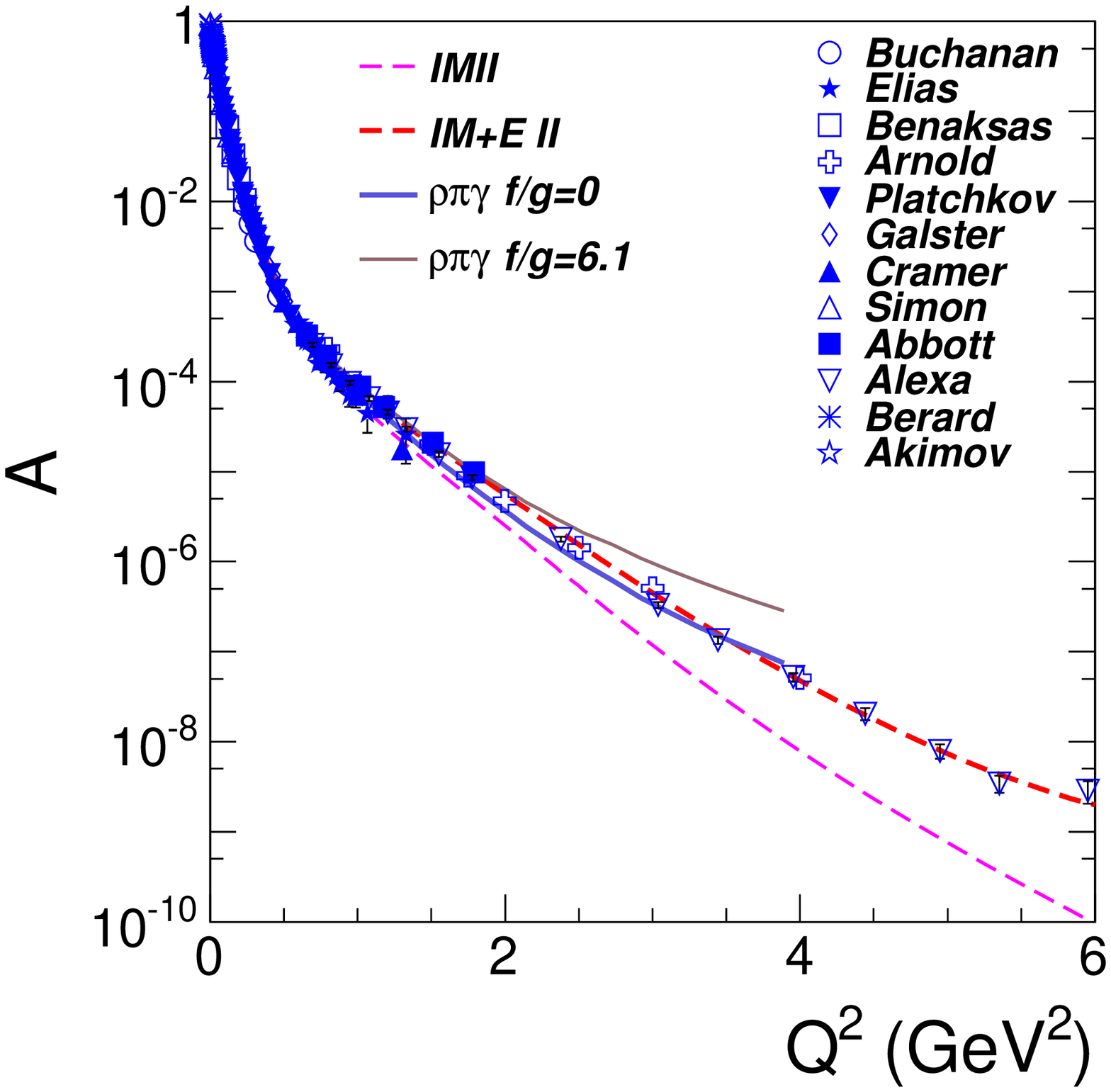}
\includegraphics[height=2.5 in, angle=0]{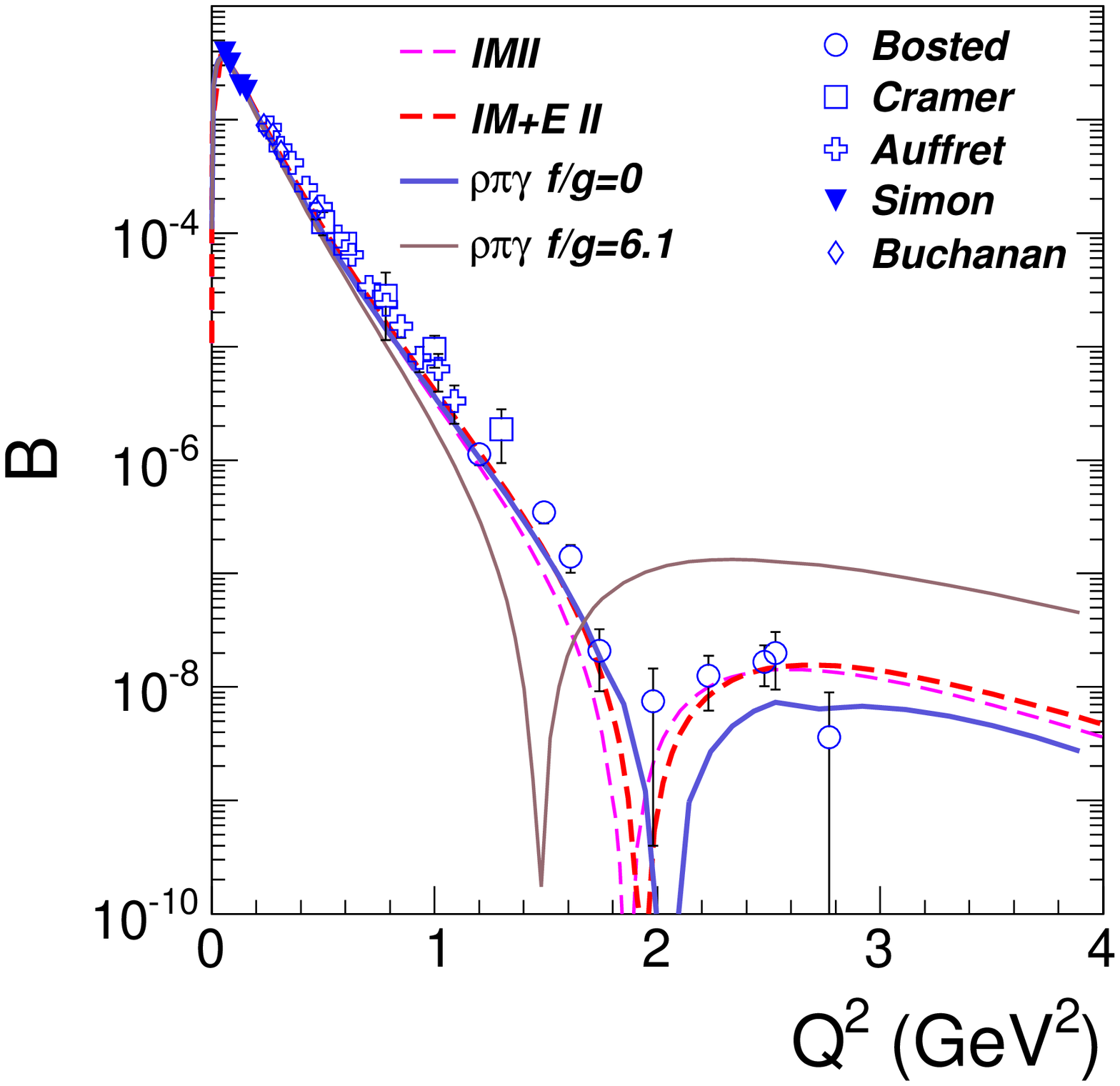}
\includegraphics[height=2.5 in, angle=0]{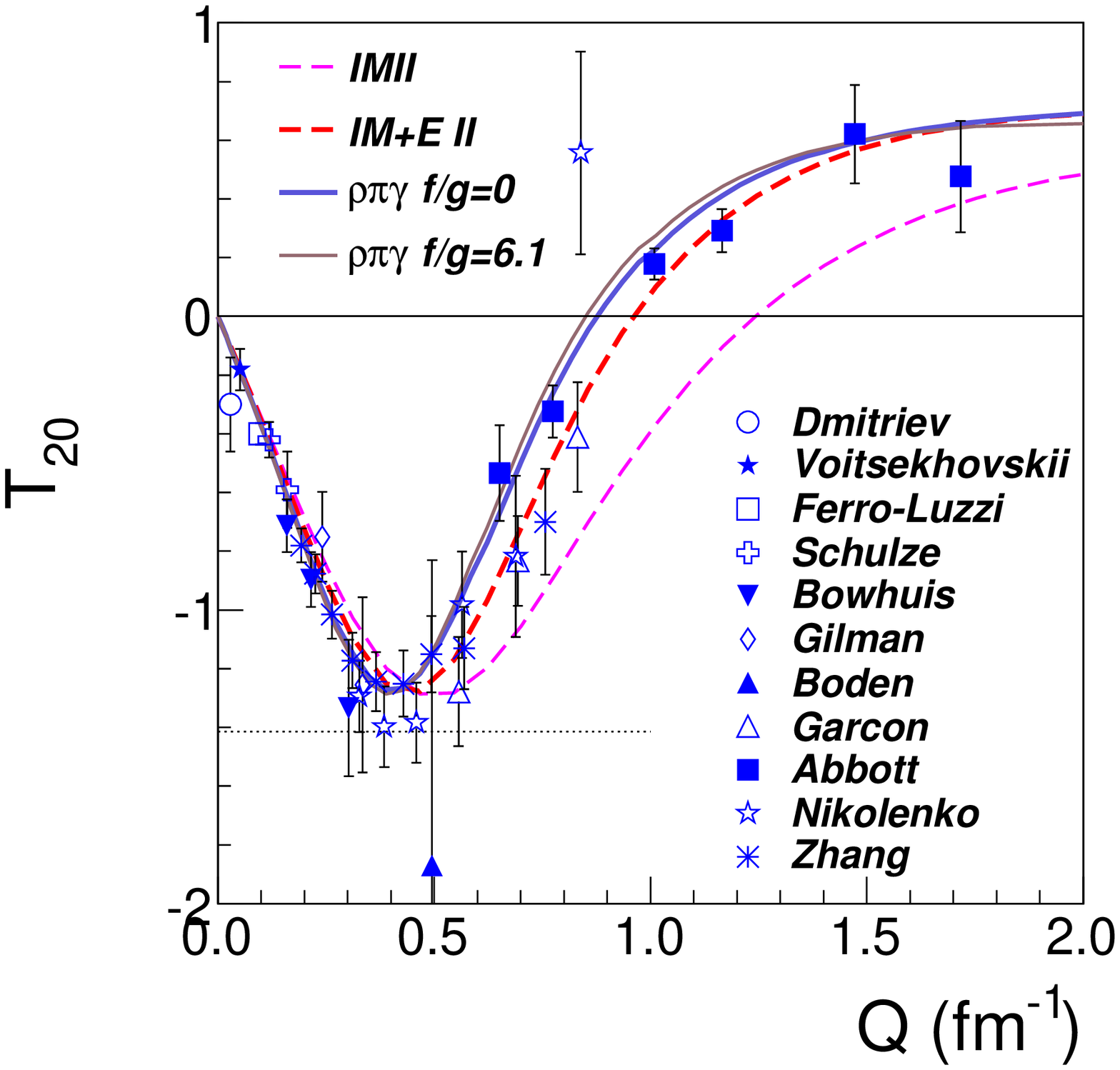}
\caption{World data for A(Q$^2$), B(Q$^2$), and t$_{20}$ = T$_{20}$ in e-d elastic scattering
compared to recent meson-nucleon calculations.
Shown are a Hamiltonian dynamics calculation \cite{Huang:2008jd} without (``IMII'')
and with (``IM+EII'') MEC, and a propagator dynamics calculation \cite{Phillips:2004nv} (``$\rho\pi\gamma$'')
with two choices (solid: f/g=0, dash: f/g=6.1) for the tensor strength of the $\rho N N$ 
interaction used in the $\rho\pi\gamma$ exchange current.
The best overall description of the data is with the ``IM+EII'' calculation.
\label{fig:abt20comb}}
\end{center}
\end{figure}

Cross sections 
\cite{Buchanan:1965zz,Elias:1969mi,PhysRevLett.13.353,PhysRev.148.1327,Arnold:1979cg,Platchkov:1989ch,Galster:1971kv,Cramer:1986kv,Simon:1981br,Abbott:1998sp,Alexa:1998fe,Berard:1974ev,Akimov:1978rk,Bosted:1989hy,Auffret:1985tg,Arnold:1986jda}
and tensor polarizations \cite{Schulze:1984ms,The:1991eg,Garcon:1993vm,Abbott:2000fg} or 
analyzing powers \cite{Dmitriev:1985us,Voitsekhovsky:1986xh,Gilman:1990vg,Boden:1990una,FerroLuzzi:1996dg,Bouwhuis:1998jj,Nikolenko:2003zq,Zhang:2011zu} 
have been measured in electron-deuteron elastic scattering. 
Since the deuteron has a spin of unity, three form factors -- charge, $G_E$, magnetic $G_M$ and quadrupole, $G_Q$ -- completely describe these observables.  The standard Rosenbluth cross section for elastic electron scattering is given by 
\begin{equation}
\frac{d\sigma}{d\Omega} =  \sigma_{Mott}\left[A(Q^2) + B(Q^2)\tan^2(\theta/2)\right],
\end{equation}
where
\begin{equation}
A = G_C^2 + \frac{2}{3}\eta G_M^2 + \frac{8}{9}\eta^2G_Q^2,
\end{equation}
\begin{equation}
B=\frac{4}{3}\eta(1 + \eta)G_M^2,
\end{equation}
and $\eta=Q^2/4M^2$ is a kinematic factor, where $M$ is the deuteron mass.
The most informative tensor polarization or analyzing power, $T_{20}$, often referred to as an alignment, is given by
\begin{equation}
T_{20} = -\frac{\frac{8}{9}\eta^2G_Q^2 + \frac{8}{3}\eta G_CG_Q + \frac{2}{3}\eta G_M^2\left[\frac{1}{2} + (1 + \eta) \tan^2(\theta/2)\right]}{\sqrt2\left[A + B\ \tan^2(\theta/2)\right]}.
\label{eq:t20}
\end{equation}
Of course, authors have pointed out that $T_{20} \rightarrow -\sqrt2$ as $Q^2 \rightarrow \infty$ and that this is a sign of the approach to pQCD scaling.  However, other estimates, discussed in the next section, indicate a more gradual approach to scaling. 
World data and two state-of-the-art calculations are shown in 
Figure~\ref{fig:abt20comb}.

Generally, the relativistic treatments of electron-deuteron scattering can be categorized \cite{Gilman:2001yh} into calculations involving Hamiltonian dynamics and those with propagator dynamics.  The former were further categorized into instant form, front form and point form by Dirac \cite{Dirac:1949cp}.
A recent informative review of Poincare invariant quantum mechanical models is given in \cite{Polyzou:2010kx}. As pointed out \cite{Gilman:2001yh}, these Hamiltonian-dynamical models suppress negative energy states and lose locality and manifest covariance. 

 Since 2000 a much better understanding of the nucleon form factors that are necessary for the calculations has become available.  
For example, the ratio of the electric to magnetic proton form factor, $G_{Ep}/G_{Mp}$, has changed dramatically compared with pre-2000 nucleon form factor extractions.  
A recent calculation \cite{Huang:2008jd} of electron-deuteron elastic scattering makes use of null plane kinematics in a Poincare invariant quantum mechanical model and also uses updated nucleon form factors \cite{Bradford:2006yz} as well as a pair-current-inspired meson exchange current (MEC).  
These calculations \cite{Huang:2008jd} of A, B and T$_{20}$ for e-d elastic 
scattering are shown in Figure~\ref{fig:abt20comb}
with curves denoted as IMII (impulse) and IM+EII (impulse + MEC).
These results indicate the importance of the MEC in the calculation.  Of course, from the discussion in Sec. 2, it is clear that at  values of $Q^2$ presently accessible in the laboratory, the approach to pQCD will not be achieved in e-d elastic scattering.  For example, one should expect to approach pQCD near 144\ GeV$^2$.  
Hence, one should expect that the relativistic N-N with MEC approach to provide a reasonable description of the existing data.  The MEC have a profound effect on A and B above 1\ GeV$^2$ and on T$_{20}$ above 0.5\ GeV$^2$.  It seems likely that MEC would tend to ``mask'' any effects from quark-gluon degrees of freedom. Furthermore, because of the small cross sections, it seems unlikely that the data can be extended to significantly higher values of momentum transfer in the foreseeable future. Nevertheless, B and T$_{20}$ have each been measured by only a single experiment at high Q$^2$ and new measurements should be performed, perhaps at Mainz or JLab, to confirm our present understanding.
While these calculations are in reasonable agreement with the data, they do not include the $\rho\pi\gamma$ MEC.  If this isoscalar MEC were to be included, it is not clear that the good agreement could be easily achieved.  Further theoretical study is necessary to determine the full effect of this MEC.  

The second main approach to relativistic electron-deuteron scattering is the propagator dynamics treatment. 
Two examples of propagator dynamics are provided by Van Orden, Devine and Gross \cite{VanOrden:1995eg} 
and Phillips, Wallace, and Devine \cite{Phillips:2004nv}.  
In the first model, the N-N interaction is described by the exchange of six mesons ($\pi$,$\eta$,$\sigma$,$\delta$,$\rho$,$\omega$). One of the nucleons is off-shell and has a form factor. This approach is often referred to as the Complete Impulse Approximation (CIA) to distinguish it from the Relativistic Impulse Approximation.  The CIA also includes two-body currents.  Recent few body calculations \cite{Pinto:2009jz} have made use of new high-precision N-N interaction models \cite{Gross:2007be,Gross:2008ps} WJC-1 and WJC-2.

The second example \cite{Phillips:2004nv} includes relativistic kinematics and the effects of negative energy states. The deuteron is described by the Bonn-B potential, a one-boson exchange model. Here the $\sigma$ meson coupling was adjusted to give the deuteron binding energy. Boosts of the two-body system and current conservation were imposed on the calculation.  In addition, isoscalar MEC were included that involve the $\gamma\pi$ contact term and the $\rho\pi\gamma$ exchange current. Relatively modern nucleon form factors were taken from the work of Kelly \cite{Kelly:2004hm}. A recent calculation of this type is compared with the data in Figure~\ref{fig:abt20comb}.

The results that have a $\rho\pi\gamma$ MEC are in reasonable agreement with the data. However, the value for the tensor strength of the $\rho N N$ interaction used in the $\rho\pi\gamma$ exchange current that vanishes gives the best agreement with the data. This value is inconsistent with $f/g = 6.1$ for the Bonn B potential.  Nevertheless, it appears possible to explain the data without invoking quark and gluonic degrees of freedom, provided that one takes some freedom with the MEC.  
A possible future direction may be to consider DSE constraints on MEC processes as indicated in \cite{Tandy:1998ha}.

\subsection{Quark-gluon approaches to the N-N interaction
and the deuteron}
\label{sec:qgdeuteron}

\begin{figure}[ht]
\begin{center}
\includegraphics[height=0.6 in]{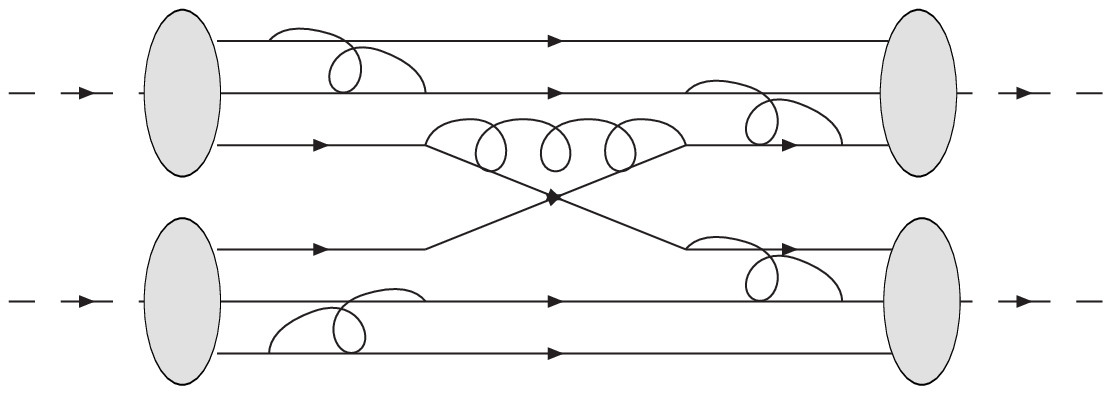}
\includegraphics[height=0.6 in]{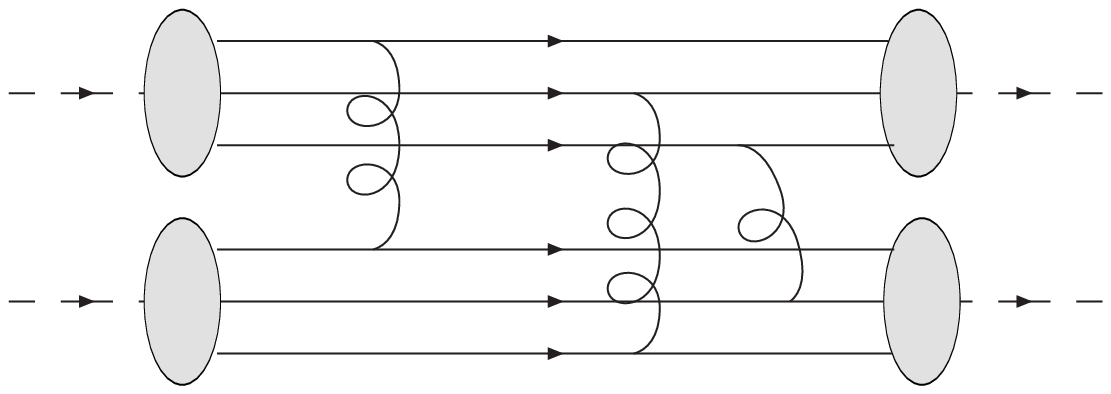}
\caption{Examples of quark diagrams of $NN$ elastic
scattering.
Left: Elastic scattering by quark interchange, with the
momentum transfer shared with other quarks by the 
exchange of five hard gluons.
Right: The independent scattering or Landshoff mechanism, 
in which each quark of a nucleon exchanges one hard gluon 
with a different quark of the other nucleon.
\label{fig:nnintpqcd}}
\end{center}
\end{figure}

The issue of quark-gluon vs. hadronic degrees of freedom
was discussed in Sec.~\ref{sec:quarkvhadronic}.
In this section we focus on the high-momentum transfer
$NN$ interaction, and the high-momentum structure of
the deuteron. It is generally accepted for these reactions
that only the leading $qqq$ Fock state of the nucleon
needs to be considered. 
As shown in \cite{White:1994tj}, high energy
hadron-hadron reactions which can proceed via
quark exchange have cross sections an order of
magnitude larger than reactions which proceed via 
gluon exchange or quark-antiquark annihilation.
This leads to the conclusion that the high-energy $NN$
reaction is dominated by quark interchange diagrams,
such as that shown in Figure~\ref{fig:nnintpqcd}.

The expected scaling for $NN$ elastic scattering is $d\sigma/dt \propto s^{-10}$,
which is approximately correct in $pp \to pp$ for $-t$ $>$ 2.5
GeV$^2$, $s$ $>$ 15 GeV$^2$
\cite{Landshoff:1973wb}.
However, the cross sections oscillate about the $s^{-10}$ 
scaling \cite{Sivers:1975dg} and there is also an 
interesting spin structure
\cite{PhysRevD.39.45,PhysRevLett.65.3241}. 
The leading explanations for these observations have been 
the interference between the pQCD and Landshoff diagrams 
\cite{PhysRevLett.61.1823} shown in Figure~\ref{fig:nnintpqcd},
or between the pQCD amplitude and
broad heavy quark resonances just above strangeness and
charm thresholds \cite{PhysRevLett.60.1924}.
A recent discussion is given in \cite{Dutta:2004fw}.
Thus, while the $NN$ interaction might have an important
perturbative quark-exchange component, there clearly
are other important contributions.

The quark counting rules lead to the helicity-conserving deuteron
form factor scaling as $1/Q^{10}$.
The only pQCD calculation of the absolute form factor
\cite{PhysRevLett.74.650} found a magnitude at least 1000
smaller than existing data, indicating either the dominance of
nonperturbative physics or of non-nucleonic, perhaps hidden-colour, 
configurations in the deuteron.

\begin{figure}[ht]
\begin{center}
\includegraphics[height=0.6 in]{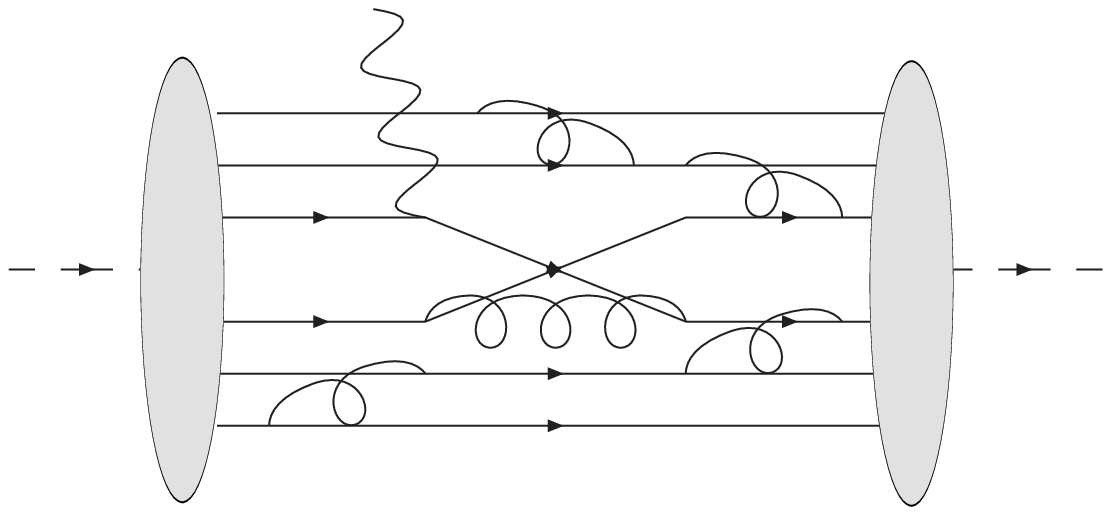}
\includegraphics[height=0.6 in]{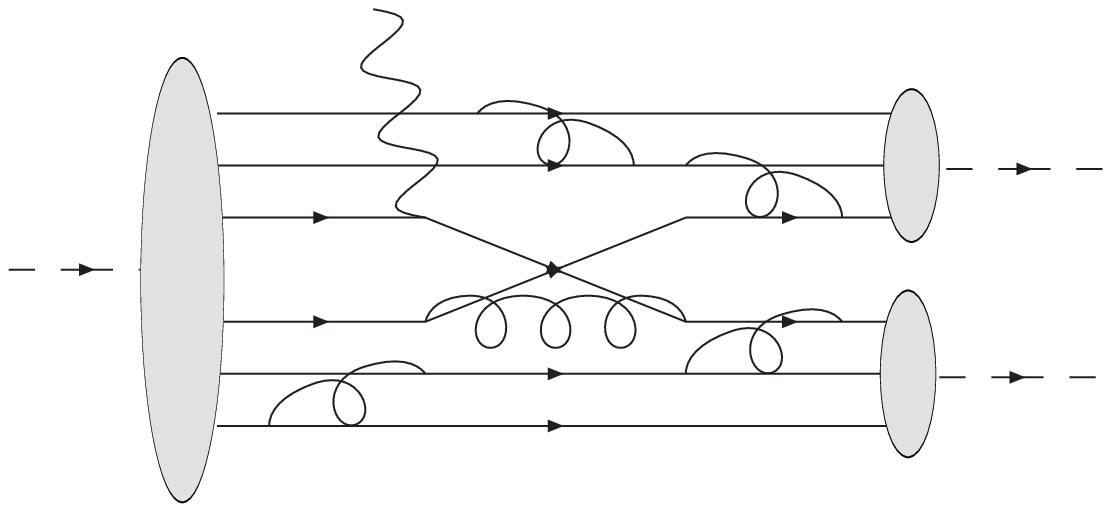}
\caption{Left: Example of a quark exchange diagram 
for the deuteron form factor.
Right: Example of a quark exchange diagram 
for deuteron photodisintegration.
In each case, the momentum is shared 
between the struck quark and the other quarks 
through the exchange of five hard gluons.
\label{fig:qediag}}
\end{center}
\end{figure}

\begin{figure}[ht]
\begin{center}
\includegraphics[height=2.2 in]{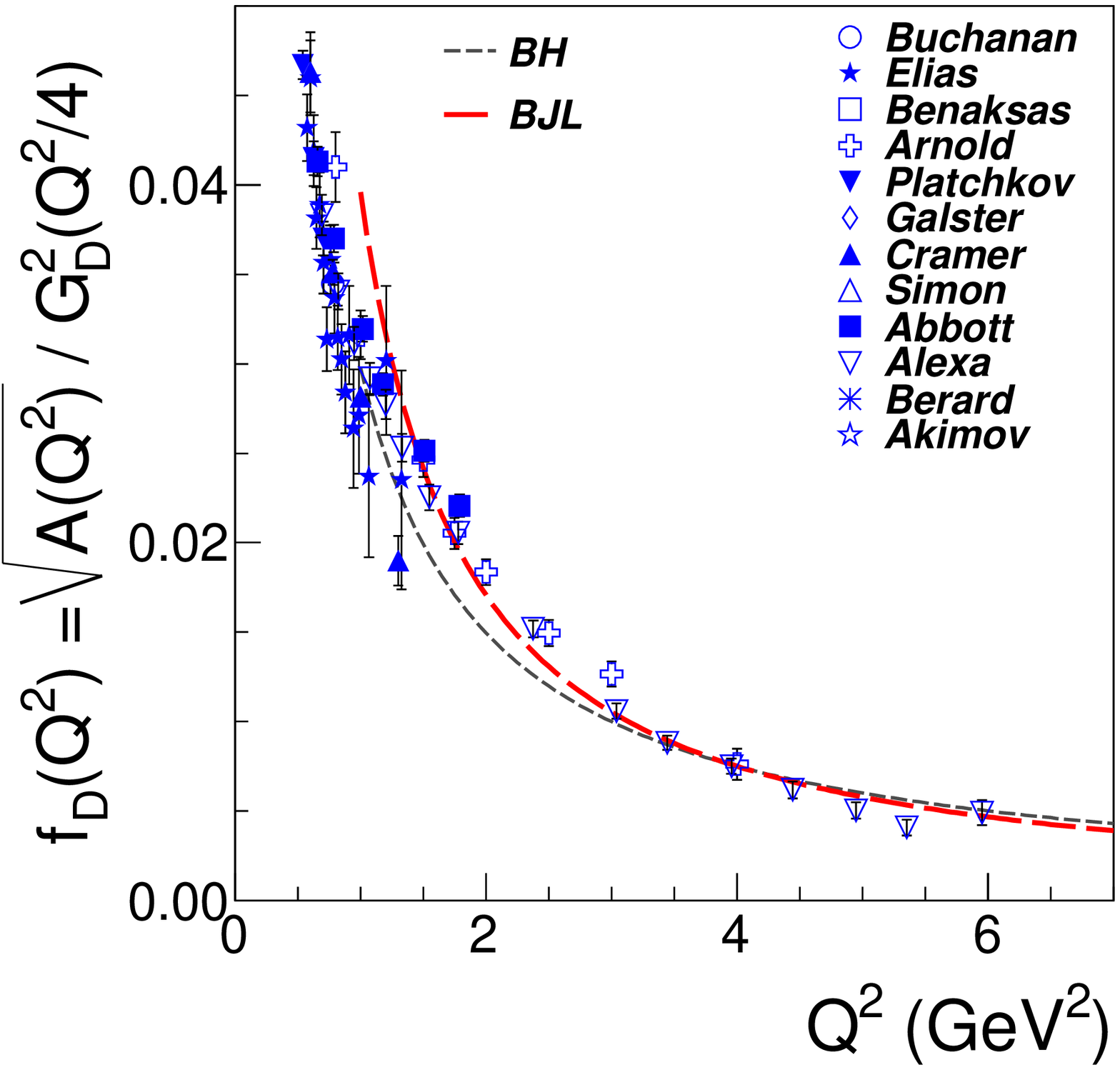}
\includegraphics[height=2.2 in]{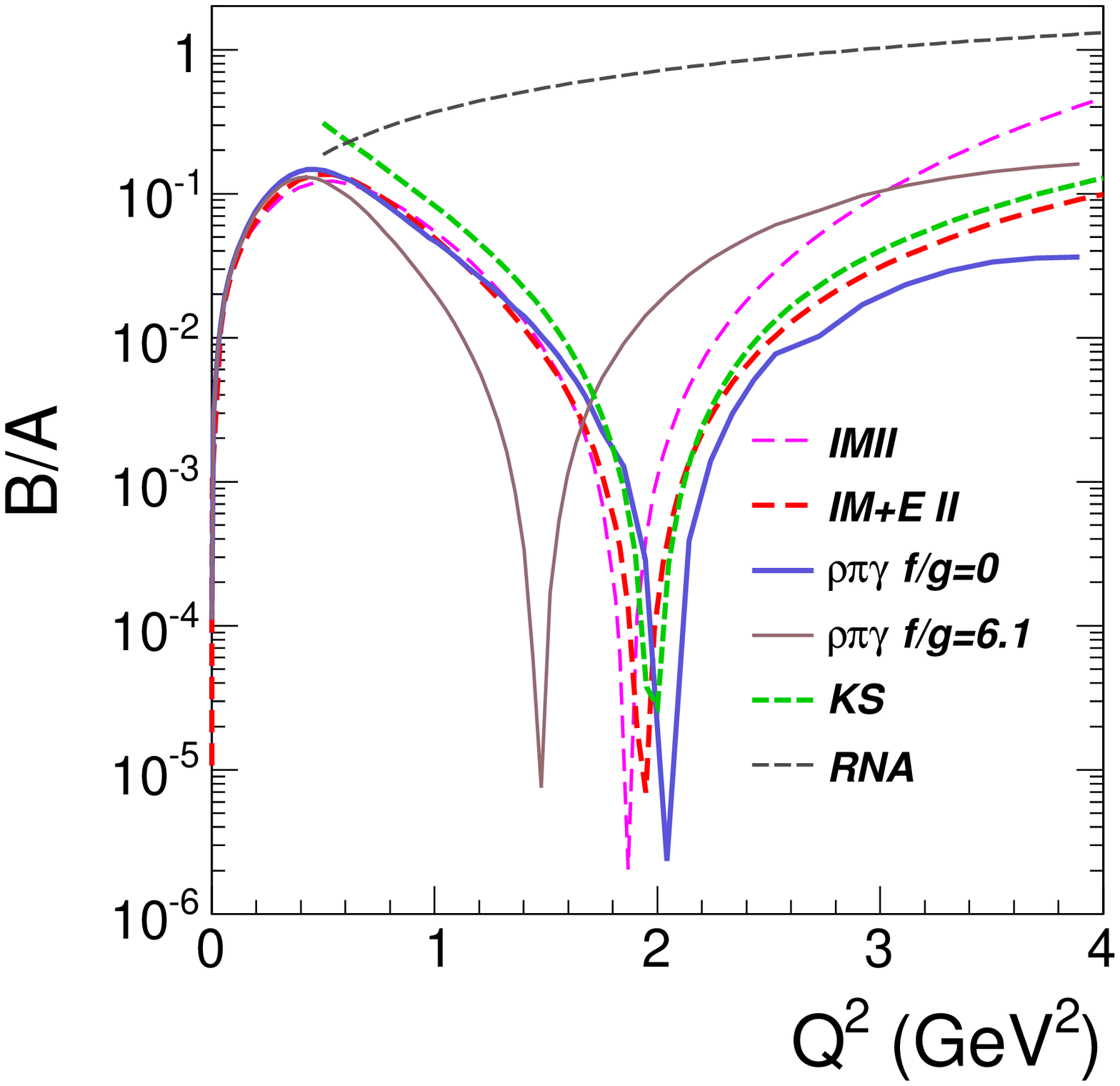}
\includegraphics[height=2.2 in]{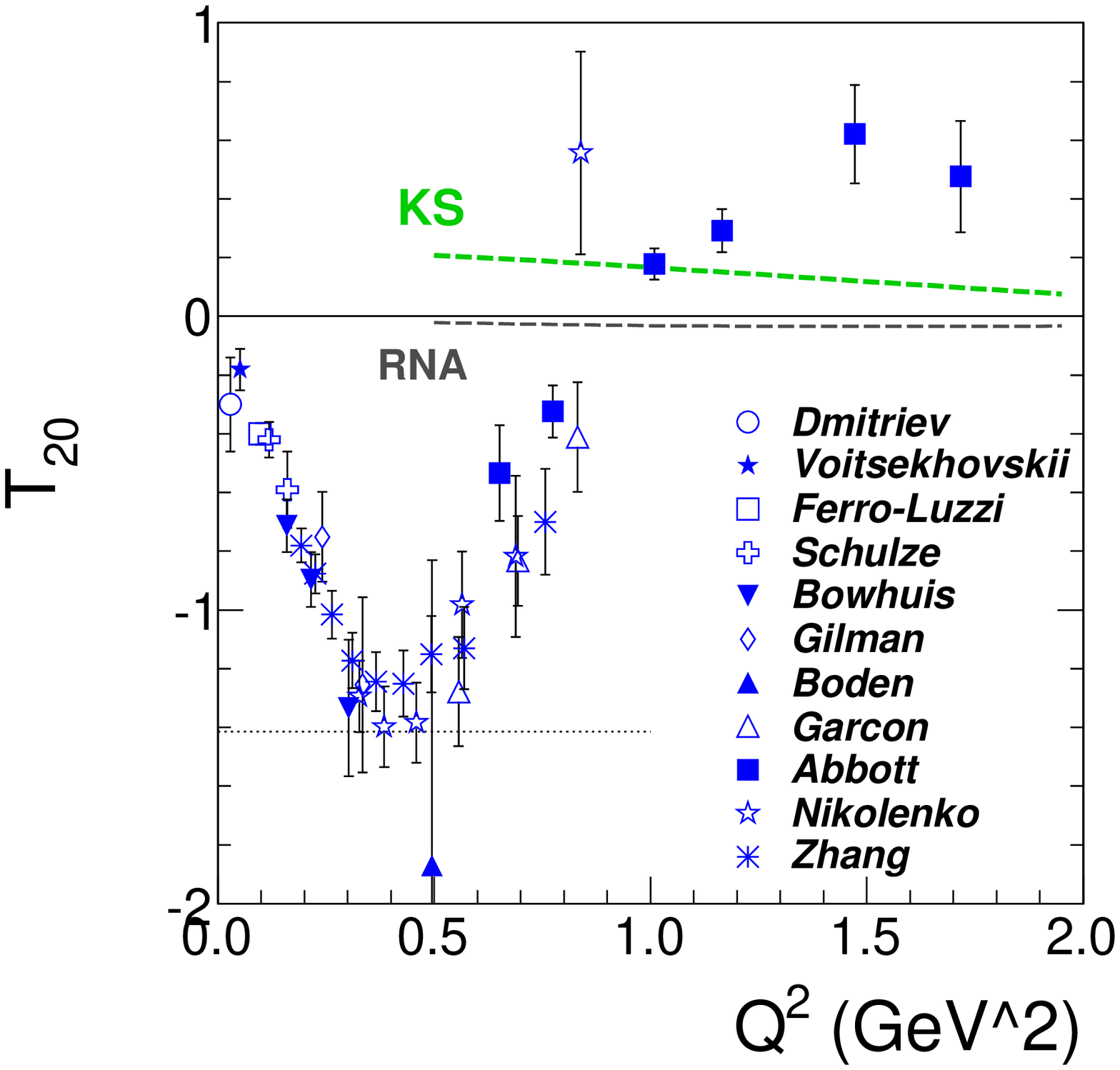}
\caption{Perturbative QCD based estimates for the 
reduced deuteron form factor $f_d(Q^2)$, the ratio of 
structure functions B/A, and the polarization $T_{20}$.
For B/A, we compare asymptotic estimates to conventional calculations.
The ``RNA'' estimate is from \cite{PhysRevD.46.2141}.
The ``KS'' estimate is from \cite{PhysRevD.49.1637}, using $Q_0^2$ = 1.15
GeV$^2$.
The ``$\rho\pi\gamma$'' calculations use propagator dynamics with different
estimates of the $\rho\pi\gamma$ meson-exchange current
\cite{Phillips:2004nv}.
The ``IMII'' and ``IM+EII'' conventional calculations use Hamiltonian dynamics \cite{Huang:2008jd}.
The ``IM+EII'' calculation is closest to the data, as shown in Figure~\ref{fig:abt20comb}.
\label{fig:pqcddff}}
\end{center}
\end{figure}

Building on the observations of \cite{White:1994tj}, one
can speculate that the deuteron form factor and
deuteron photodisintegration reaction are dominated by
quark-exchange diagrams such as those shown in Figure~\ref{fig:qediag}.
This is the underlying picture originally adopted for these
reactions in the reduced nuclear amplitudes (RNA) approach
\cite{Brodsky:1983kb}, which works surprisingly well for the
helicity conserving
deuteron form factor -- here extracted from the A structure function -- 
to quite low $Q^2$, as shown in Figure~\ref{fig:pqcddff}.
The reduced form factor $f_D(Q^2)$ was estimated to be a monopole
in \cite{Brodsky:1983kb}; the ``BH'' line shown uses
$(1+Q^2/m_0^2)^{-1}$ with $m_0$ = 0.1 GeV.
Subsequently \cite{Brodsky:1983vf} estimated that $f_D(Q^2)$
should vary logarithmically with $Q^2$ as 
$(\ln(Q^2/\Lambda^2))^{-1-(2/5)C_F/\beta}/Q^2$;
the ``BJL'' line shown uses 
$\Lambda$ = 0.1 GeV, $C_F$ = 4/3, and $\beta$ = 29/3. 
The hard rescattering model discussed further in Sec.~\ref{sec:dgp}
can be viewed as a further refinement of this approach applied to 
high-energy photodisintegration, and is the most successful 
existing explanation of that reaction.

Scaling arguments from pQCD have also been applied to various combinations
of the deuteron form factors.
Carlson and Gross \cite{PhysRevLett.53.127}, based on helicity-flips leading to an
extra power of $Q^2$ in the falloff of form factors, estimated
that $G_M$ and $G_Q$ fall as $Q^{-12}$ and $G_C / G_Q = 2 \eta/3$
where $\eta = Q^2 / 4 M_d^2$.
Subsequently, Brodsky and Hiller \cite{PhysRevD.46.2141}
found the asymptotic ratio of deuteron form factors 
to be $G_C$ : $G_M$ : $G_Q$ = $1-2\eta/3$ : 2 : -1.
Kobushkin and Syamtomov \cite{PhysRevD.49.1637}  extended this
by including the subleading helicity-flip form factors.
The interference between helicity non-flip and helicity flip form
factors allows $B/A$ is reproduced down to $\approx$ 1 GeV$^2$,
including the minimum -- see Figure~\ref{fig:pqcddff}.
But the calculation does not reproduce $T_{20}$ well,
even though it crosses over the data at $\approx$ 1 GeV$^2$
as can also be seen by in Figure~\ref{fig:pqcddff}.
Cao and Wu \cite{PhysRevC.55.2191} found
the asymptotic ratio of form factors to be 
$G_C$ : $G_M$ : $G_Q$ = 
$1+{8 \over 3}f+{2 \over 3} c f^2 - {2 \over 3}\eta $ : $2(1+f)$ : -1.
Here $f$ is a parameter determined by the sign change in $G_M$, 
$c$ is a constant of order unit, and $c^{\prime}$ is another constant 
of order unity that does not appear in the asymptotic form factor ratio.
This approach leads to results similar to \cite{PhysRevD.49.1637}.
These issues are reviewed in \cite{Carlson:1997bs}.

The preceding approaches are all based on a perturbative picture of
the deuteron.
There have also been some efforts at nonpertubative quark models 
of the $NN$ system.
Maltman and Isgur \cite{PhysRevD.29.952} found that the 6-quark
system strongly clusters into an $NN$ configuration.
De Forest and Mulders \cite{DeForest:1986xf} in a simple model
examined the effects of antisymmetrization of quarks in the two 
nucleons. They concluded that antisymmetrization breaks the
concept of factorization, such as that suggested by
\cite{Brodsky:1983kb}, and becomes increasingly important
with increasing momentum.
Dijk and Bakker \cite{Dijk:1989ia} studied the deuteron within the
quark-compound bag model. The basic philosophy is that the $A=2$
system has a short-range 6-quark component and a long-range
$NN$ component.
As shown in \cite{Gilman:2001yh}, the approach yields a good
description of deuteron form factors, comparable to the best
conventional relativistic $NN$ models.
Robson \cite{PhysRevC.61.015202} treats the deuteron as a sum of
two 3-quark harmonic oscillator systems, with quark orbits in the
different nucleons required to be orthogonal. The model gives a
semiquantitative description of data; it includes a quark-correlation 
effect which improves the description, with  similar effects to the 
$\rho\pi\gamma$ meson-exchange term in conventional models.
A recent estimate \cite{Pirner:2010fw} of the effects of 
6-, 9-, $\ldots$ quark bags on nuclear structure indicated 
for example that these structures could account for
quasifree electron scattering data at $x > 1$ which
is more traditionally interpreted as indications of short-range
nucleon correlations in nuclei -- see \cite{Higinbotham:2011aa}
for a discussion of recent experimental results in this area.

In summary, the best approaches to understanding the deuteron
structure remain relativistic hadronic models tied to the underlying
$NN$ force, despite some uncertainties in this approach.
QCD-inspired models have some success. Estimates more firmly
based on QCD fail particularly for $T_{20}$.

\subsection{Photodisintegration of the deuteron}
\label{sec:dgp}

The most recent review of deuteron photodisintegration 
remains \cite{Gilman:2001yh}, where a fuller discussion
can be found of physics presented here.
The photodisintegration reaction provides large center-of-mass 
energy $W$ for incident photon energies of a few GeV. 
In the hadronic picture, hundreds of resonance channels 
would be potentially excited by $E_{\gamma}$ = 4 GeV, which 
would be natural to sum over in a reaction model with quark-gluon 
degrees of freedom, just as in deep inelastic scattering.
Also similar to DIS, at GeV energies and large angles there is
GeV scale four-momentum transfer $-t$ and $-u$, or equivalently 
transverse momentum $p_T$, again suggesting that quark models
are appropriate for understanding the reaction dynamics.

Several approaches to the underlying quark dynamics have been developed.
A simple pQCD approach
\cite{Matveev:1972gb,Brodsky:1973kr,Brodsky:1974vy} 
predicts that cross sections follow the constituent counting rules, 
$d\sigma/dt$ $\propto$ $s^{-11}$,
and polarizations are constrained by hadron helicity conservation, e.g.,
$p_y$ = $C_{x'}$ = $C_{z'}$ = 0. In fact, cross sections follow the
constituent counting rules better than they should, but polarizations do not
follow hadron helicity conservation (HHC) \cite{Gilman:2001yh}.
The failure of HHC is no longer surprising.
In the DSE approach, massive quarks lead to HHC violating couplings.
Furthermore, the importance of quark orbital angular momentum in
the structure of the nucleon is now widely appreciated.
The RNA approach \cite{Brodsky:1983kb} attempted to extend the validity
of the pQCD $s$ dependence to lower energies by including expected 
threshold kinematic factors, but the simple $s^{-11}$ dependence
actually agrees better with the data in the $E_{\gamma}$ $>$ 1 GeV
region, as shown below in Sec.~\ref{sec:g3heppn}.

The dominance of quark-interchange diagrams in the $NN$ interaction, discussed in
Sec.~\ref{sec:qgdeuteron}, leads to the Hard Rescattering Model (HRM).
The HRM is based on the photon
being absorbed by a pair of quarks being exchanged between the two
nucleons, and relates photodisintegration to $NN$ scattering. 
Because $NN$ data roughly follow the counting rules, photodisintegration should as well.
(This general idea was investigated within a different physical 
model in \cite{JuliaDiaz:2002ra}.)

The dominance of planar diagrams in QCD \cite{'tHooft:1974hx} leads to
the quark gluon string model (QGS). Deuteron photodisintegration is treated as
3-quark exchange, and modelled with nonlinear
Regge trajectories, which have been used to describe a number of
high-energy reactions, to photodisintegration.

These quark models have provided some insight into the underlying
dynamics, along with semi-quantitative predictions of cross sections 
and polarizations, but the $\gamma d \to pn$ data were insufficient
at the time of \cite{Gilman:2001yh} to uniquely identify the
underlying dynamics. Since that review, there have been several
advances.

Grishina {\it et al.} \cite{Grishina:2002ph,Grishina:2003as}
realized that the pQCD limit for the linearly polarized photon
asymmetry, $\Sigma(\theta_{cm} = 90^{\circ})$ $\to$ -1, was due to the
assumption of isoscalar photon coupling. For isovector photon
coupling, the limit becomes +1. The $\Sigma$ asymmetry data
at 90$^{\circ}$ are all positive above about 600 MeV, and hint at 
an increase with energy above 1 GeV.

The CLAS collaboration \cite{Mirazita:2004rb} measured a complete 
set of angular distributions for $E_{\gamma}$ =
0.5 -- 3 GeV and a center of mass angle range as much as
10 -- 160$^{\circ}$; these data agree with and dramatically extend 
earlier angular distribution measurements \cite{Schulte:2002tx}.
The CLAS data demonstrated \cite{Rossi:2004qm} that the 
threshold for the scaling behaviour is given approximately by 
$p_T$ = 1.1 GeV/$c$, confirming the observation of a $p_T$ threshold,
based on a much smaller data set \cite{Schulte:2001se}.

Tensor polarization asymmetries in deuteron photodisintegration
were measured \cite{Rachek:2006dz} for $E_{\gamma}$ $\approx$
70 -- 500 MeV. These data are generally well predicted by modern
hadronic theory \cite{Schwamb:2010zz,Schwamb:2000zr}, though detailed differences exist.
\begin{figure}[ht]
\begin{center}
\includegraphics[height=3.0 in, angle=0]{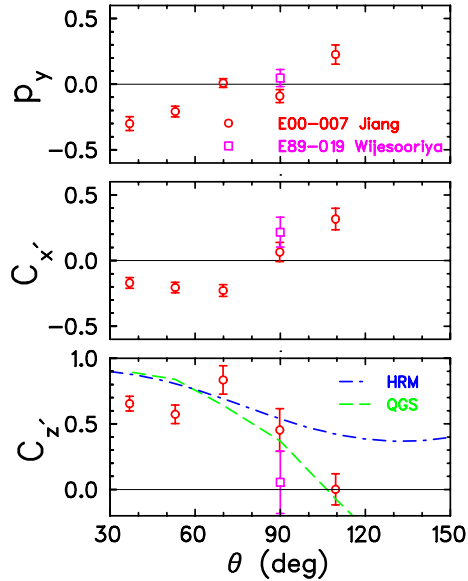}
\caption{Angular distributions for recoil polarization at $E_{\gamma}$
  $\approx$ 2 GeV, adapted from Jiang \etal \cite{Jiang:2007ge}. 
Calculations are from HRM \cite{Sargsian:2003sz} and QGS \cite{Grishina:2002ph,Grishina:2003as}.
The data from E89-019 Wijesooriya \cite{Wijesooriya:2001yu}
are at  $E_{\gamma}$  $\approx$ 1.9 GeV.
Adapted from \cite{Jiang:2007ge}. 
\label{fig:dgp-pol-ad}}
\end{center}
\end{figure}

An angular distribution of recoil polarizations was measured
at $E_{\gamma}$ $\approx$ 2 GeV \cite{Jiang:2007ge} -- see
Figure~\ref{fig:dgp-pol-ad}. 
The induced polarization and transverse transferred polarization
vary with angle so that they cross zero near $\theta_{cm}$ =
90$^{\circ}$. In the HRM \cite{Sargsian:2003sz}, a natural explanation
is that with isovector dominance these polarization components are 
proportional to the $NN$ amplitude $\phi_5$ that vanishes at 90$^{\circ}$.
The longitudinal transferred polarization is large
at forward angles and tends to falls with angle. This behaviour
qualitatively agrees with the predictions shown.

Khokhlov, Knyr and Neudatchin \cite{PhysRevC.75.064001}
studied photodisintegration for $E_{\gamma}$ = 1.1 - 2.3 GeV, 
using a point-form relativistic quantum mechanics approach 
with an optical potential derived from $NN$ elastic scattering data
up to 3 GeV.
Their calculated cross sections reproduced the data well, but there
are no published calculations of polarization observables.

Recoil polarizations were measured for $E_{\gamma}$ $\approx$
280 - 360 MeV \cite{Glister:2010ft}. 
This is the region in which the induced proton polarization starts to 
dramatically diverge from calculations. The behaviour was confirmed 
in a finely binned systematic data set; polarization transfers were
also determined. 
Detailed differences were seen with the best modern hadronic 
calculations \cite{Schwamb:2010zz,Schwamb:2000zr}.

In summary, while conventional hadronic models provide the best description
of low-energy deuteron photodisintegration, it is difficult to extend
these models into the GeV region. QCD-inspired models tied to the
$NN$ force, the HRM and QGS models, provide both a semi-quantitative
description of cross sections and a qualitative description of some
polarization observables. 
The new high-energy polarization data in particular test 
models of the underlying dynamics, but the similarity in predictions
prevents identifying a correct model.
While the focus has now largely turned towards $^3$He disintegration as a means
of understanding the physics, as discussed in Sec.~\ref{sec:g3heppn},
there is interest \cite{Duttadeltadelta:2011} in testing the 
recent calculation of \cite{Granados:2010cj} of enhanced
$\Delta\Delta$ pair production by disintegrating the short-range, 
6-quark structure of the deuteron.

\section{Photoreactions in the light nuclei}
\label{sec:phtoreactionsagt2}

Similar to the case for the deuteron, light $A$ = 3, 4 nuclei
have been studied through elastic scattering.
We will not consider the elastic form factors in any detail as, 
similar to the deuteron case, data can be well
explained by conventional nuclear theory with 
meson-exchange currents, but do not go to high $Q^2$.
Published elastic $^3$He and $^3$H form factor data, e.g.,
\cite{Amroun:1994qj,PhysRevLett.86.5446}, extend only
up to $\approx$1.5 GeV$^2$, while published $^4$He 
data \cite{Arnold:1978qs} are limited to about 2 GeV$^2$;
see \cite{Sick:2001rh} for a review.
Unpublished data have been taken by the Hall A
collaboration up to $\approx$3.5 GeV$^2$ \cite{Gomez:2004}.

\label{sec:g3heppn}

High-energy photodisintegration is most studied for $^3$He.
High-energy photodisintegration of $^3$He leads to
$pp+n_{spectator}$, $pn+p_{spectator}$, and three-body final states.
The basic idea for the $\gamma ^3{\rm He} \to pp + n_{spectator}$ reaction
\cite{Brodsky:2003ip} is to compare hard $pp$ disintegration 
from $^3$He with hard $pn$ disintegration from the deuteron.
Models not able to predict the absolute cross sections might still
be able to predict the ratio of these two processes.
The initial predictions were for the cross sections for $\gamma pp$ 
cross sections to be similar to or larger than those of $\gamma pn$;
this contrasts with low energies where the  $\gamma pp$ cross sections
are an order of magnitude smaller than those for $\gamma pn$,
which is explained by the vanishing magnetic dipole moment for two protons
coupled to spin 0.
It was also expected in the HRM that due to the observed oscillation
in the $pp$ elastic cross section that $\gamma pp$ cross section
would also exhibit oscillations.
The HRM theory was further developed in
\cite{Sargsian:2008zm,Granados:2010gn}.

The $n$ spectator actually provides some advantages compared to the
$\gamma d \to p n$ case.
In the impulse approximation, the variation in initial-state
neutron momentum varies the $\gamma pp$ center of mass energy,
so that the energy dependence of the reaction can be measured in
a single setting.
The neutron light-cone momentum fraction,
$\alpha_n = (E_n-p_{zn})/M$, is nearly unaffected by soft
final-state rescatterings, and thus is sensitive to
the neutron's wave function --  if the $\gamma pp$
disintegration is a short distance process, this implies
large $pp$ momentum in the initial state, which through
correlations in the wave function leads to high neutron 
momentum, and a harder $\alpha_n$ distribution. 
The opposite is true if the  $\gamma pp$ process
depends on long-range processes.

\begin{figure}[ht]
\begin{center}
\includegraphics[height=2.5 in, angle=0]{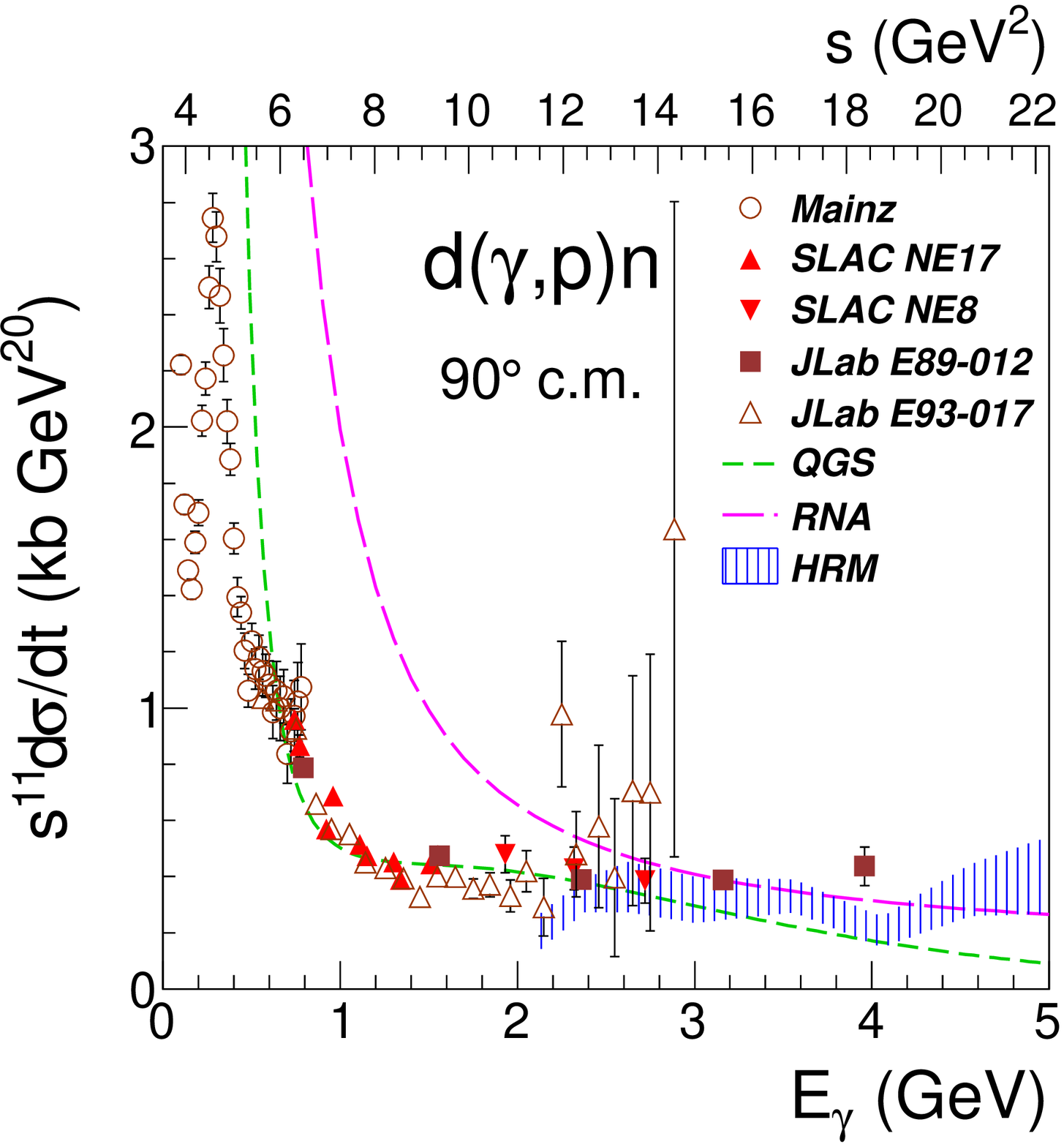}
\includegraphics[height=2.5 in, angle=0]{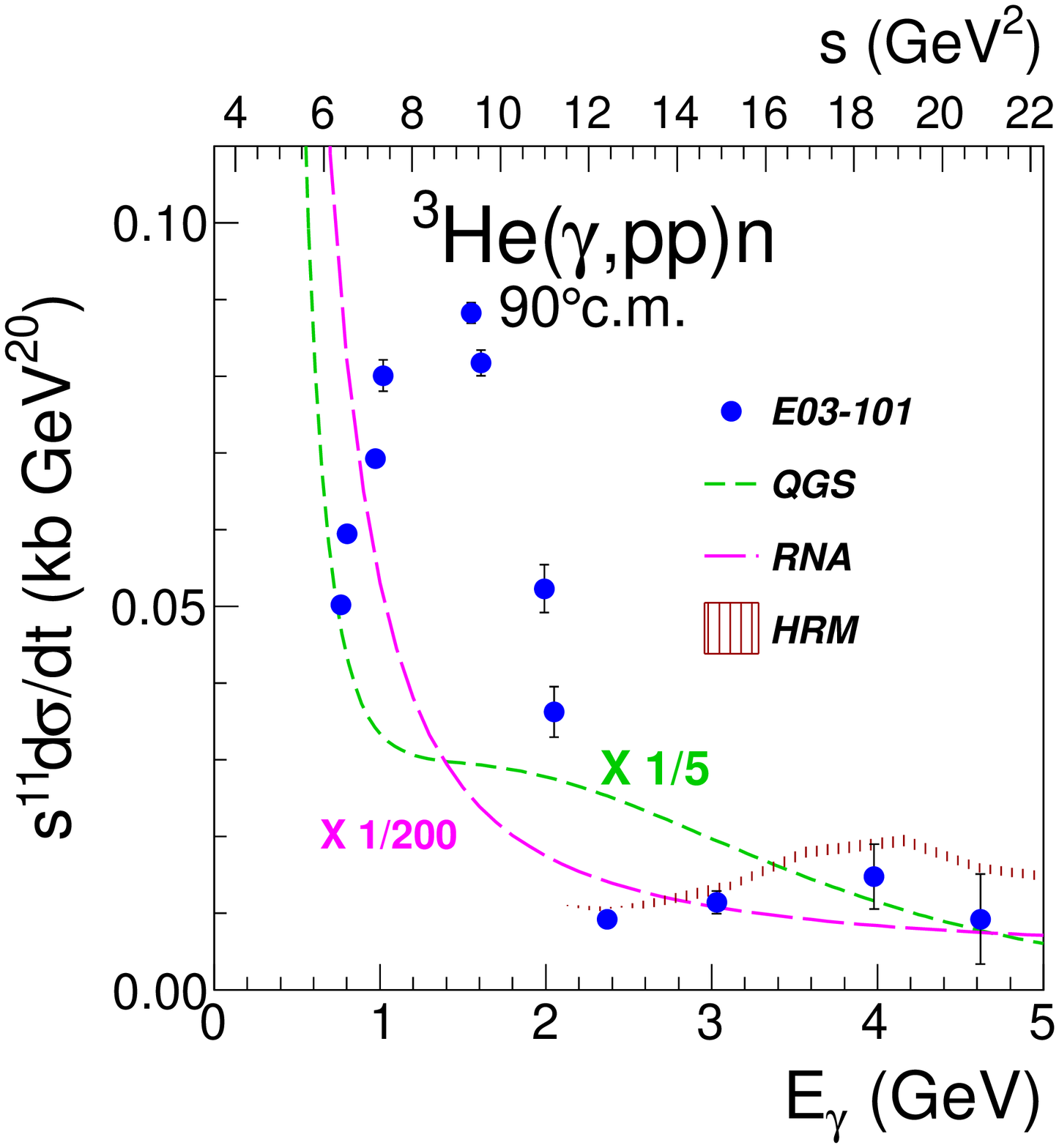}
\caption{Left: cross sections for $\gamma d \to pn$ at $\theta_{cm}$ =
  90$^{\circ}$, showing the high-energy $s^{-11}$ scaling.
Calculations are labelled
QGS \cite{Kondratyuk:1993wh},
RNA \cite{Brodsky:1983kb} and
HRM \cite{Frankfurt:1999ik}.
Data are labelled
Mainz \cite{Crawford:1996ka},
NE8 \cite{Napolitano:1988uu,Freedman:1993nt},
NE17 \cite{Belz:1995ge},
E89-012 \cite{Bochna:1998ca} and
E93-017 \cite{Mirazita:2004rb}.
Right: cross section for $\gamma ^3{\rm He} \to pp+n_{spectator}$,
showing the high-energy $s^{-11}$ scaling and the much smaller
size of the absolute cross sections.
The RNA and QGS curves have been adjusted by the given factor from
their expected magnitudes given in \cite{Brodsky:2003ip} to be of 
similar size to the data.
The HRM calculations are from \cite{Sargsian:2008zm,Granados:2010gn}.
The cross sections shown have an experimental cut $|p_n|$ $<$ 
100 MeV/$c$; the estimated correction factor for higher neutron
momenta is a factor of two.
Adapted from \cite{Pomerantz:2009sb}.
\label{fig:g3he2ppn}}
\end{center}
\end{figure}

Figure~\ref{fig:g3he2ppn} shows the only published set of
high-energy $\gamma ^3{\rm He} \to pp + n_{spectator}$ data
\cite{Pomerantz:2009sb}.
The results can be divided into two energy regions.
For 1 GeV $<$ $E_{\gamma}$ $<$ 2 GeV there is a several
hundred MeV wide region with a peak or peaks in the 
$\theta_{cm}$ = 90$^{\circ}$ cross sections.
At the peak the cross sections are slightly less than 1/2
of the $\gamma pn$ cross sections.
The origin of this peak is unclear; speculations include
three-body processes or resonance excitation -- though it
should be remembered that there is no indication of resonance
excitation in the $\gamma d \to pn$ data in this energy range.
For $E_{\gamma}$ $>$ 2 GeV the cross sections exhibit
approximate $s^{-11}$ scaling, at a level a factor of 20 smaller
than the $\gamma d \to pn$ data.
It is important to note that the scaling is indeed the $s^{-11}$ 
of a two-body process and not the $s^{-17}$ of a three-body 
process.
The idea of a neutron spectator that does not affect
the scaling is supported by the data.
The small size of the cross sections was unexpected,
and prevented determining the $\alpha_n$ distribution
adequately, or whether the fall off is slightly slower than $s^{-11}$
as expected in the HRM.
The small size is now understood in the HRM to arise from
a cancellation between two $NN$ amplitudes due to opposite 
signs, which was not recognized in
\cite{Brodsky:2003ip}. It is not known at this time whether or 
not there will be similar effects in the other approaches to the
quark dynamics.

Other high-energy photodisintegration experiments include
$\gamma ^3{\rm He} \to ppn, \; pp+n_{spectator}$
\cite{Niccolai:2004ne} and $\gamma ^4{\rm He} \to pt$
\cite{Nasseripour:2009in}, which extend only up to 
$E_{\gamma}$ $\approx$1.5 GeV, not into the scaling region.
The $\gamma ^3{\rm He} \to pd$ channel has been measured
in both JLab Hall A and CLAS,
apparently into the scaling region, but is unpublished
\cite{PomerantzIlieva:2011}.

\section{Perspectives}

QCD, proposed more than three decades ago, is the accepted theory of the strong interaction.  Nevertheless, the application of QCD to reactions with light nuclei remains elusive.  Electromagnetic interactions with hadrons and light nuclei provide the most sensitive test for QCD effects in nuclei since the electromagnetic interaction is relatively well known and calculations can be performed for the simplest systems. Indeed, calculations with perturbative QCD can be performed, however, these calculations have routinely underestimated the exclusive cross section data at accessible energies.  Models involving a factorization process where the incoming high energy photon interacts perturbatively with a quark, but subsequent interactions are relatively soft have had some degree of success. The future theoretical developments likely lie with nonperturbative approaches such as Dyson-Schwinger equations or lattice QCD.  It is essential for experiment to map out the long-range behaviour of QCD. Future facilities such as the upgraded CEBAF at Jefferson Lab, COMPASS-II at CERN, Drell-Yan experiments at FNAL, J-PARC and RHIC, as well as a possible future electron ion collider hold promise to provide illuminating data for our simplest processes where QCD can be applied. 

Although new results for the neutral pion transition form factor from Belle call the puzzling BaBar observations into question, a confirmation of these findings for one of our most elementary processes will be necessary.
The form factors for the pion and nucleons will be pushed to significantly higher momentum transfers in the coming decade.  These results provide a sensitive determination of the role of dynamical chiral symmetry breaking on the structure of hadrons. While the EMC effect has taught us that the momentum distribution of quarks in bound nucleons is significantly different from those in free nucleons, we do not yet have information on the quark flavor dependence of the EMC effect.  The stage is being set to perform new measurements that will reveal this substructure of the EMC effect.  The very idea of medium modifications remains controversial,
and experimental studies of exclusive reactions probing this area remain inconclusive.
While we have had good evidence for a color transparency effect in meson electroproduction, new data from the upgraded CEBAF are necessary to be convincing. While new exclusive photoreaction experiments can be performed at higher energy, the most interesting might be exclusive photopion production since it involves a relatively small number of constituents.
The short-range behavior of the deuteron remains a mystery after decades of experimentation and theoretical development. There is only one set of measurements of the magnetic form factor and the tensor polarization in electron-deuteron elastic scattering at high momentum transfer.  The magnetic form factor seems to be best described by calculations with either no or an incomplete $\rho\pi\gamma$ meson exchange correction.  It should be straightforward to provide new measurements of the magnetic form factor in future experiments.

\section*{Acknowledgments}
We especially thank C. D. Roberts, I. C. Cloet, D. Phillips,
S. Wallace and W. Polyzou for providing tables of their calculations,
and D. Dutta, M. Paolone and I. Pomerantz for their help in preparing
several of the figures in this work.
We also heartily thank K. Hafidi, S. Pieper, W. Polyzou, and C. D. Roberts,
for extremely useful discussions.
This work was supported by: 
Department of Energy, Office of Nuclear Physics, contract no.\ DE-AC02-06CH11357 for Argonne National Laboratory and the U.S.\ National Science Foundation grant PHY 09-69239 for Rutgers University.

\section*{References}
\bibliography{htoq_bib}

\end{document}